\documentclass[12pt,a4paper]{article}
\usepackage{amssymb,amsmath,graphicx,a4wide}
\newcommand{\avg}[1]{\langle #1\rangle}
\newcommand{\abs}[1]{\lvert #1\rvert}
\newcommand{\kopt}{k_{\mathrm{opt}}}
\newcommand{\Xopt}{X_{\mathrm{opt}}}
\newcommand{\amin}{\alpha_{\mathrm{min}}}
\newcommand{\ee}{\mathrm{e}}
\newcommand{\dd}{\mathrm{d}}
\begin{document}
\vspace*{16pt}
\begin{center}
{\LARGE Market Model with Heterogeneous Buyers}\\[10pt]
\large
Mat\' u\v s Medo and Yi-Cheng Zhang\\[10pt]
Physics Department, University of Fribourg, CH-1700 Fribourg
\end{center}
\vspace{16pt}

\begin{abstract}
In market modeling, one often treats buyers as a~homogeneous
group. In this paper we consider buyers with heterogeneous
preferences and products available in many variants. Such a
framework allows us to successfully model various market
phenomena. In particular, we investigate how is the vendor's
behavior influenced by the amount of available information and
by the presence of correlations in the system.\\[6pt]
\end{abstract}

\section{Introduction}
The standard economics textbooks make the supply-demand law as
one of the pillars of the modern economic theory. However, many
people, especially economists (see for example \cite{stiglitz}),
gradually realize that the most important factor is missing in
the traditional supply-demand law. The study of complex
systems~\cite{SantaFe,minority-book,martino-marsili} has already
led to novel approaches to market phenomena. In a previous
work~\cite{supply-demand}, one of us introduced a simple
framework to treat both quality and information capability,
yielding a generalized supply demand law. However, in the
previous paper, product is simply characterized by a single
scalar variable: quality. In the modern economy we face a much
more complex world, where products have many attributes and
consumers have heterogeneous tastes~\cite{happier-world}. These
preferences cannot be simply represented as price and quality
alone. We therefore generalize the previous work to allow
multiple variants of each product as well as many different
tastes among consumers.

Thus the producers face a dilemma: whether to target the average
taste by producing a single or a few variants to leverage the
economy of scale, or to match precisely each consumer's
taste~\cite{Bakos}. We shall see that the answer depends on the
information level that the producers may access: whether they
know, and how well they know the consumers' preferences. In
addition, producers face also the nonlinear production costs.
All these factors have to compromise to yield a combined result
that gives various degrees of product diversity. With our
approach, the supply demand problem of producers with the
capability of producing variations and consumers' diverse tastes
becomes a matching
problem~\cite{stable-marriage,stable-marriage-new}, where many
mathematical and statistical mechanic tools are available to
handle the complexity of the combinatorial problem.




In this paper we build a market model and investigate its
behavior under various circumstances. In the first part of the
paper we do not consider correlations between preferences of the
parties included in the system. While unrealistic, this
assumption allows us to discover basic properties of the model
and outline the way of reasoning which can be used also in
later, more realistic considerations. In the second part of the
paper we discuss correlations and the ways how they can be
introduced to the system. The last part of the paper deals with
the consequences of the correlations for the model.

\section{General framework: one vendor with many buyers}
Let's start with a market where only one vendor and $M$ buyers
are present. The vendor can produce $N$ different variants of
a product (e.g. many different shoes). With regard to the
market, he has to decide which variants it is optimal to
produce. We assume that all buyers satisfied with the offer buy
one item, others stay out of the trading. Buyers in the market
we label with lowercase Latin letters ($i=1,\dots,M$). The
different variants the vendor can produce we label with Greek
letters ($\alpha=1,\dots,N$). The price of variant $\alpha$ we
label as $P_{\alpha}$. We assume that every variant can be
produced in as many pieces as it is needed and as fast as it is
needed.

The simple structure sketched above offers us enough space to
model basic features of real markets. To establish
a~mathematical model for the market we have to introduce some
assumptions about participants' preferences and their
consequences on the trading process. To keep complexity of the
model at minimum we assume that buyer's opinion about a variant
can by represented by one scalar quantity, which we call
\emph{cost} and label it with $x$; we assume $x\in[0;1]$. The
smaller is the cost $x_{i,\alpha}$, the bigger is the
probability that buyer $i$ is satisfied with variant $\alpha$
when asked. Preferences of the vendor are easier to introduce;
they are represented by costs which he suffers during production
and sale of a particular variant. The cost for variant $\alpha$
we label $y_{\alpha}$ and after a proper rescaling of monetary
units  $y_{\alpha}\in[0;1]$. To simplify our considerations, we
arrange the variants in order of cost: $y_1<y_2<\dots<y_N$.

We stress a conceptual difference between vendor's and buyer's
costs. The seller's cost $y_{\alpha}$ is strictly monetary---it
represents a real amount of money (although in arbitrary units).
In contrast, the buyer's cost $x_{i,\alpha}$ has no tangible
interpretation, it simply represents something as airy as
happiness with the given variant.

The vendor is able to produce $N$ different variants. However,
when he is producing more variants, his expenses grows due to
need of an additional investment. The vendor's tendency to
produce only few different variants can be modeled e. g. by a
nonlinearity of expenses (doubled production of one single
variant does not require doubled expenses). We adopt another
approach; we assume that to initiate the production of a
variant, the vendor has to pay additional charge $Z>0$ (this
represents initial costs).

Now let's assume that the vendor offered $k$ most favorable
variants (thus $\alpha=1,\dots,k$, $k\leq N$) to customers and
the number of units sold of variant $\alpha$ is $n_{\alpha}$.
The total vendor's profit is
\begin{equation}
\label{vendors-profit}
X(k,\{n_{\alpha}\})=\sum_{\alpha=1}^k
n_{\alpha}(P_{\alpha}-y_{\alpha})-kZ.
\end{equation}
Here the last term $kZ$ comes for the initial costs of $k$
produced variants, $P_{\alpha}-y_{\alpha}$ is the profit for one
sold unit of variant $\alpha$. Due to the monetary rescaling
used to confine $y_{\alpha}$ to the range $[0;1]$, units for
profit, initial costs and prices are arbitrary.

It is natural to assume that when buyer $i$ is asked about
interest to buy variant $\alpha$, the decision is based
on the cost $x_{i,\alpha}$. We formalize this by the assumption
that the probability of acceptance is a function of the variant
cost; this function we call acceptance function. Obviously,
$f(x)$ is a decreasing function of the cost $x$. Moreover, we
assume $f(0)=1$. This means that if a buyer considers a variant
to be the perfect one, she surely buys it.

When we offer a random variant to one buyer, the acceptance
probability is
\begin{equation}
\label{average-acceptance}
\int_0^1\pi(x)f(x)\dd x\equiv p.
\end{equation}
Here $\pi(x)$ is the probability distribution of cost $x$
(i.e.~it defines what ``to offer a random variant'' really
means). The probability $p$ of accepting a random proposal is an
important parameter of the model. From our everyday life we know
that largely we do not agree to such an offer. For this reason
we assume $p\ll1$ in our calculations.

\begin{figure}
\centering
\includegraphics[scale=1]{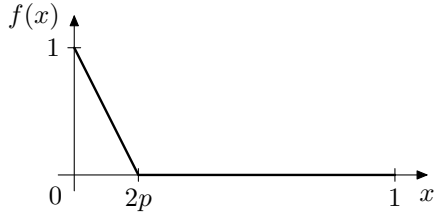}
\caption{One particularly simple choice for the buyers'
acceptance function $f(x)$.}
\label{fig-fx}
\end{figure}

One example of a reasonable choice for the acceptance function
is (see fig.~\ref{fig-fx})
\begin{equation}
\label{acceptance-function}
f(x;p)=
\begin{cases}
1-x/2p & (0\leq x\leq 2p),\\
0 & (2p<x).
\end{cases}
\end{equation}
with $p<0.5$. This choice is especially convenient due to its
simplicity. If we now assume the uniform distribution of the
buyer's costs, $\pi(x)=1$ for $0\leq x\leq 1$, parameter $p$ of
the acceptance function (\ref{acceptance-function}) is just the
probability $p$ of accepting a random offer introduced in the
previous paragraph.

In the rest of this paper we assume that the prices of all
variants are the same and equal to $1$, $P_{\alpha}=1$. This
relieves us from many technicalities, and helps to highlight
important features of the model. Nevertheless, generalization to
various prices is straightforward.

\section{No correlations in costs}
We begin our investigation with the simplest case of the
presented model---the market without correlations, where all
costs $y_{\alpha}$ and $x_{i,\alpha}$ are mutually independent.
We model this by costs uniformly distributed in the range
$[0;1]$. To keep variants ordered, we first draw their costs
and then we renumber all variants achieve $y_1<y_2<\dots<y_N$.
It follows that after averaging over realizations, the formula
$\avg{y_{\alpha}}=\alpha/(N+1)$ holds.

\subsection{A vendor without knowledge of buyers' preferences}
\label{sec-noinfo}
If a vendor wants to discover which variants are most
acceptable for buyers, in a market without correlations each
buyer has to be asked for preferences. This cannot be done in
big markets, thus it is natural to investigate the case with no
information about buyers' preferences on the vendor's side.
In sec.~\ref{sec-informed} we show that even an expensive global
opinion survey brings only a negligible contribution to the
vendor's income.

Without any information about preferences, the vendor is not
able to discover which variants are most favored by buyers.
Therefore the best strategy is to offer variants that are most
favorable from his point of view. Let's label the number of
variants the vendor is willing to offer as $k$. We assume that
all these variants are available to buyers simultaneously,
similarly to different types of shoes available in a shoe shop.
Every buyer goes through the offered variants and decides
whether some of them are suitable or not.

From the buyer's point of view, the vendor makes random
proposals; the probability of accepting one particular offer is
thus by definition equal to $p$. The probability $P_A$ that one
particular buyer accepts one of $k$ proposed variants is
complementary to the probability $(1-p)^k$ of denying all
offered variants. Thus we have
\begin{equation}
\label{PA}
P_A=1-(1-p)^k\approx 1-\ee^{-pk},
\end{equation}
where the approximation used is valid for $pk\ll 1$, i.e. for
very choosy consumers (than $p$ is a small quantity) and a small
number of offered variants. Now the average number of items sold
by the vendor to all $M$ buyers is $MP_A$. Since no correlations
are present, the average number of items sold of variant
$\alpha$ is $\avg{n_{\alpha}}=MP_A/k$, it is a decreasing
function of $k$.

\begin{figure}
\centering
\includegraphics[scale=1]{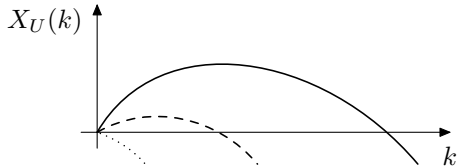}
\caption{Expected profit of the vendor without informations,
$X_U(k)$, drawn against $k$ for small value of initial costs
(solid line), medium initial costs (dashed line) and high
initial costs (dotted line). In the last case the condition
$Z>Mp$ is fulfilled and the optimal vendor's strategy is to stop
the production.}
\label{fig-profit}
\end{figure}

The quantity of vendor's interest is the total profit $X$
introduced in (\ref{vendors-profit}). Its expected value can be
found using $\avg{n_{\alpha}}$, $\avg{y_{\alpha}}$, and $P_A$. We
obtain
\begin{equation}
\label{profit}
X_U(k)=
M\big(1-\ee^{-pk}\big)\left(1-\frac{1+k}{2(N+1)}\right)-kZ.
\end{equation}
Here the subscript $U$ reminds that we are dealing with an
``uninformed'' vendor. This function is sketched in
fig.~\ref{fig-profit} for three different choices of initial
cost $Z$. The optimal number of variants the vendor should offer
maximizes his profit. One can easily show that when $X_U'(0)<0$,
$X_U(k)<0$ for all $k>0$. Thus the condition $X_U'(0)<0$, which
can be rewritten as $Z>Mp$, characterizes a market where the
optimal vendor's strategy is to stop the production and stay
idle.

Since for every product numerous numerous variations can be
made, the total number of variants the vendor can offer, $N$, is
large. Thus we are allowed to assume that the optimal number of
offered variants satisfies the condition $\kopt\ll N$ and solve
the maximization condition $X_U'(k)=0$ approximately. We obtain
\begin{equation}
\label{noinfo-opt}
\kopt=\frac1p\,\ln\frac{Mp}{Z},\quad
\Xopt=M-\frac Zp\bigg(1+\ln\frac{Mp}Z\bigg),
\end{equation}
where $\Xopt$ is the optimal expected profit,
$\Xopt=X_U(\kopt)$. The used approximations are valid when
$Z\gg M/N$ and $p\ll1$.
\begin{figure}
\centering
\includegraphics[scale=0.26]{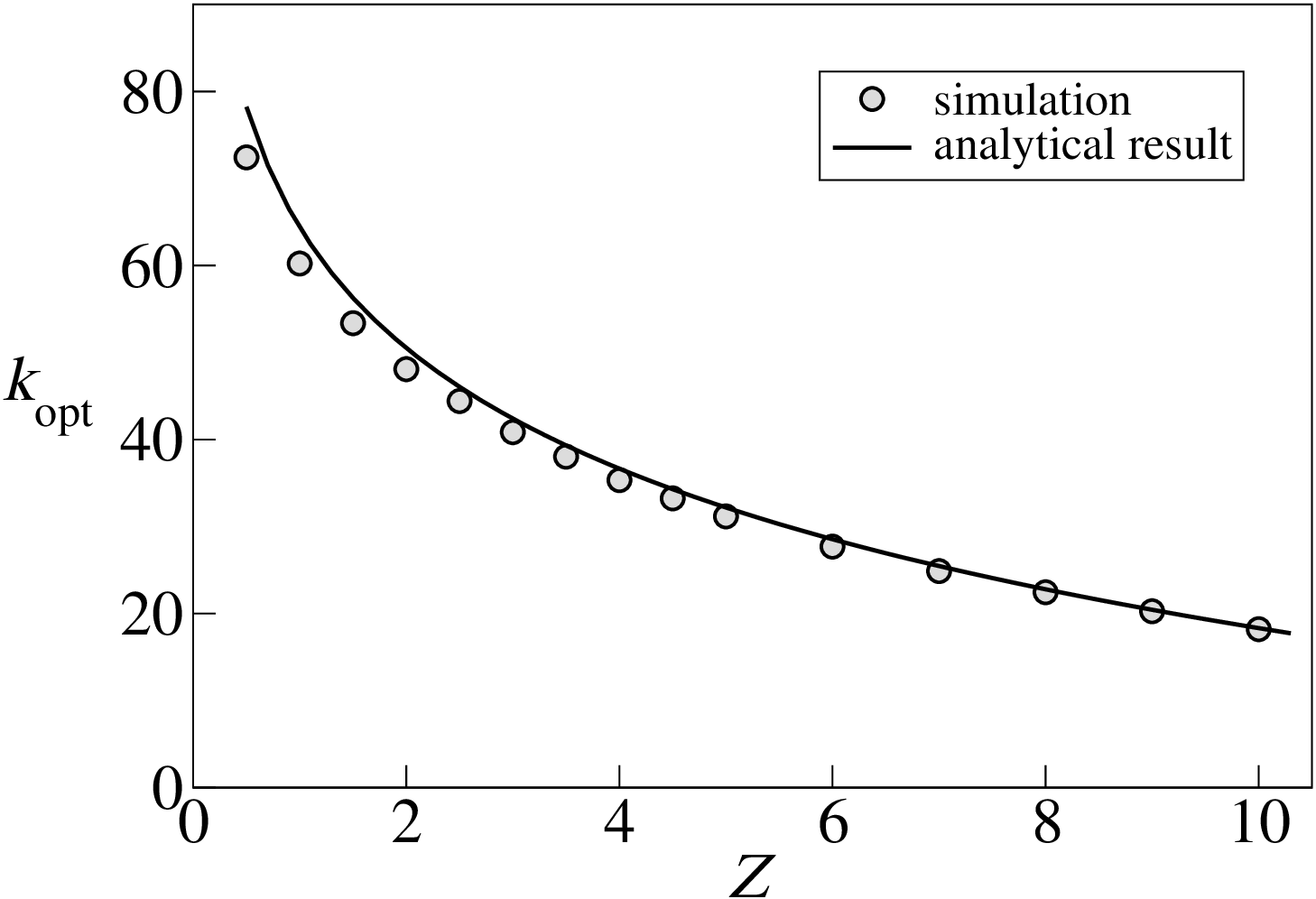}\quad
\includegraphics[scale=0.26]{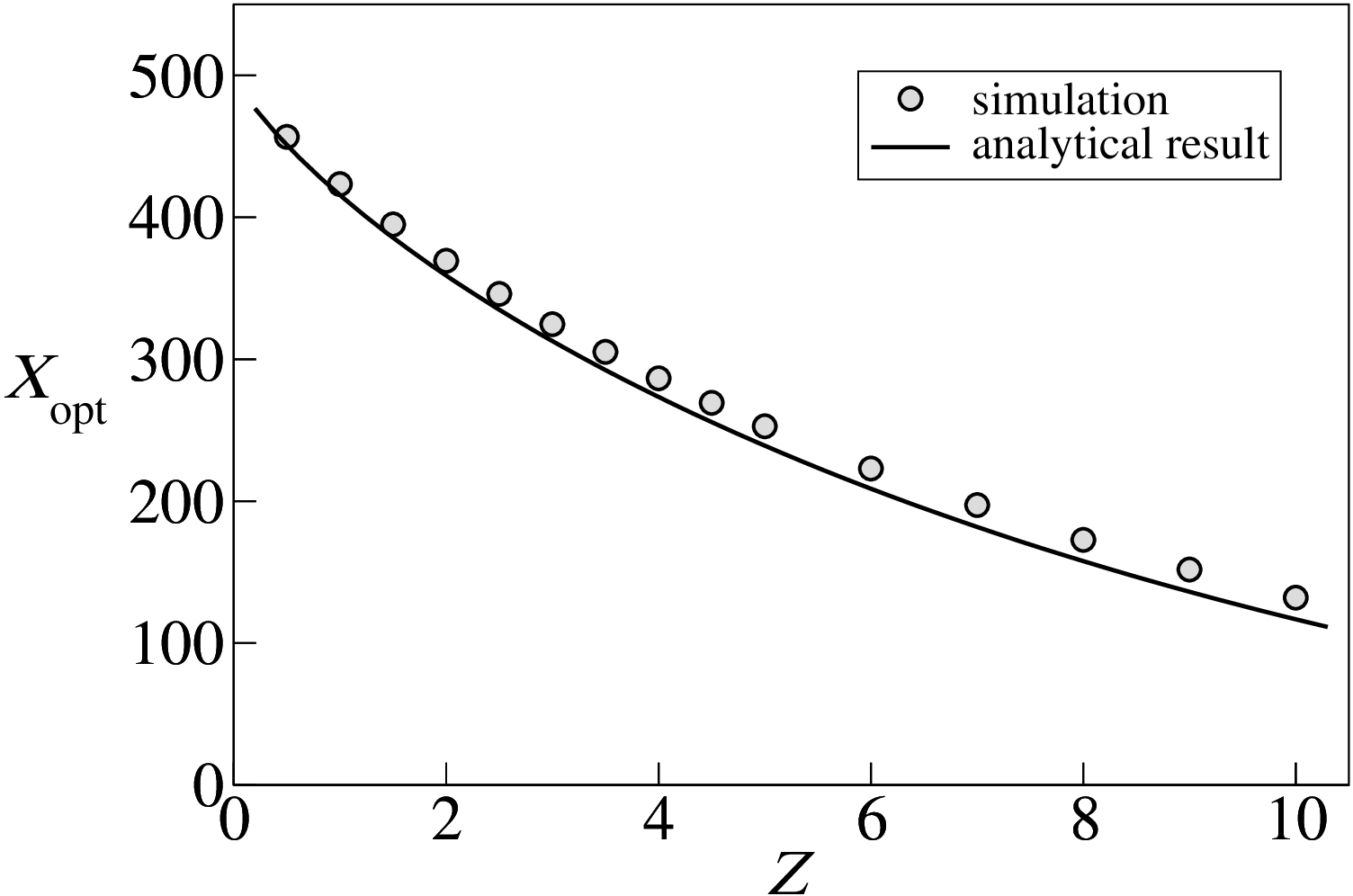}
\caption{The optimal number of offered variants and the optimal
profit as functions of initial cost $Z$ for $M=500$, $N=2\,000$,
and $p=0.05$. Numerical results (empty circles) are averages
over 1\,000 realizations, analytical results (solid lines) come
from eq.~(\ref{noinfo-opt}). For the optimal profit arbitrary
units are used.}
\label{fig-kopt}
\end{figure}

\begin{figure}
\centering
\includegraphics[scale=0.45]{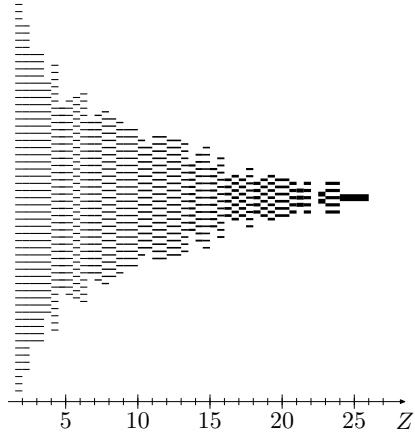}
\caption{Differentiation of the vendor's production for various
initial costs $Z$: single realization of the model (no averaging
present) with $M=500$, $N=2\,000$, $p=0.05$. Vertical axis has
no significant meaning, it serves purely to distinguish
different variants.}
\label{fig-differentiation}
\end{figure}

In fig.~\ref{fig-kopt}, these results are shown to match a
numerical treatment of the problem. In the figure we see how
the initial cost $Z$ influences diversity of the vendor's
production: decreasing $Z$ increases differentiation of the
vendor's supply in full agreement with expectations. We can
examine this feature in detail if we plot the optimal number of
offered variants against $Z$ for one particular realization of
the model as it is shown in fig.~\ref{fig-differentiation}
(thickness of the lines is proportional to the number of buyers
of a variant).

\subsection{Improvement of the vendor's profit by a sequential
offering of variants}
So far we dealt with a very passive approach of the vendor.
While offering $k$ variants to the market, he had no influence
on the sale. In consequence, due to the absence of correlations
in the system, every offered variant had the same average number
of items sold. In a big market this is a natural approach. While
the use of advertising can promote some variants, its treatment
exceeds our scope.

In a small market a personal offering is possible. The vendor
can promote favorable variants to increase the profit simply by
offering the most favorable variant first. If a buyer is not
interested, the second most favorable variant follows, etc.
The average sale of the first variant is $\avg{n_1'}=Mp$, for
the second variant it is $\avg{n_2'}=M(1-p)p$ and in general we
have $\avg{n_{\alpha}'}=Mp(1-p)^{\alpha-1}$. Hence the expected
total sale is
\begin{equation*}
\sum_{i=1}^k Mp(1-p)^{i-1}=M\big[1-(1-p)^k\big].
\end{equation*}
This is equal to the expected total sale $MP_A$ of the
uninformed vendor in the previous section. We can conclude that
the vendor's profit improvement (if any) does not come from
an increased total sale but rather from an increased sale of
the variants that are more profitable for the vendor.

\begin{figure}
\centering
\includegraphics[scale=0.26]{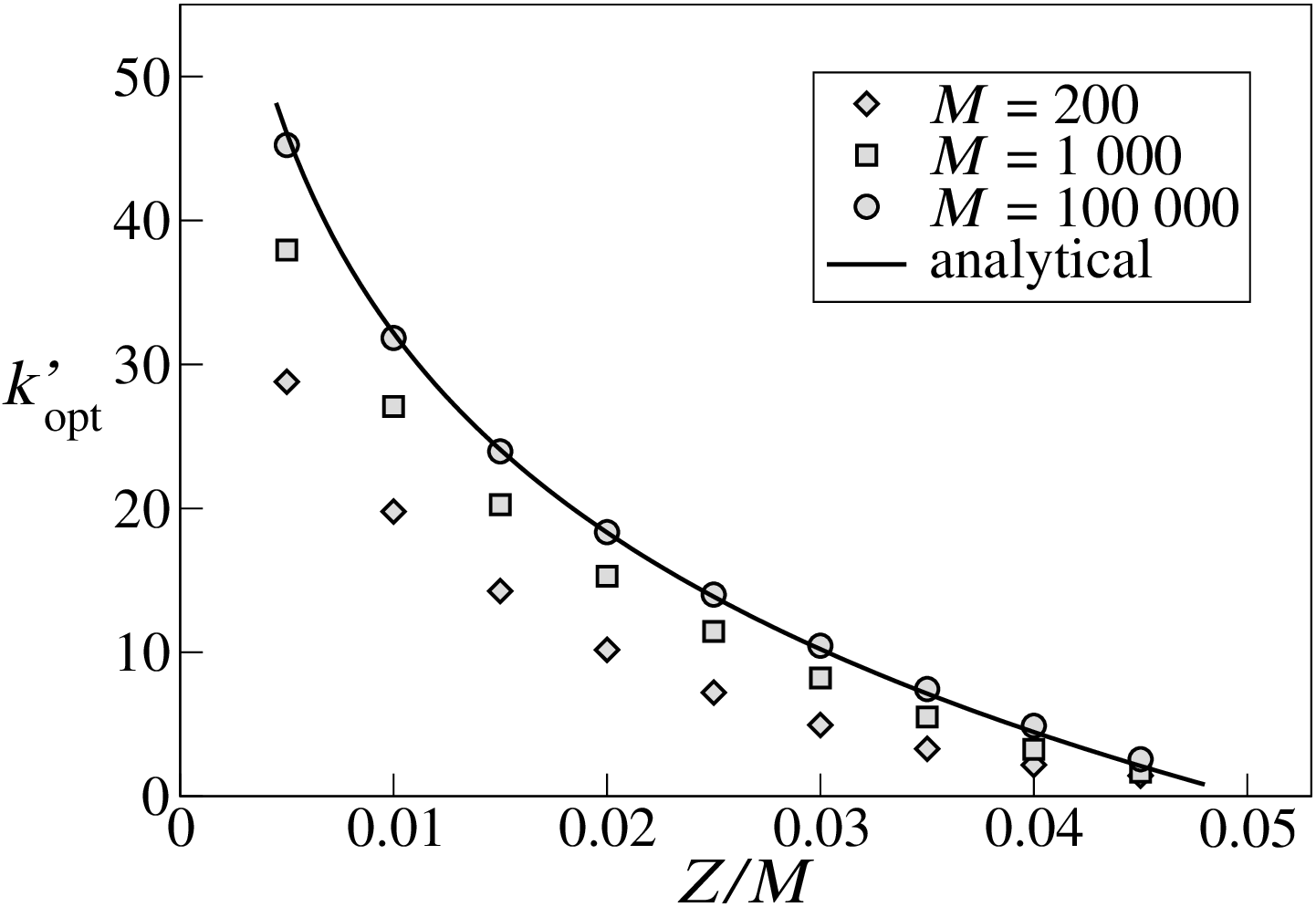}
\caption{Successive offering: numerical and analytical results
for the vendor using stopping condition described in the text
in the markets with various sizes (on horizontal axis we have
$q\equiv Z/M$). All numerical results are obtained as average of
10\,000 realizations with $p=0.05$, $N=2\,000$; solid line
represents eq.~\ref{kopt-improved}.}
\label{fig-improved}
\end{figure}

Now we investigate the optimal number of variants to offer in
this case, $\kopt'$. Since $\avg{n_{\alpha}'}$ decreases with
$\alpha$, at some moment it is not profitable to offer one more
variant and the vendor's profit is maximized. The corresponding
equation $\avg{n_{\alpha}'}=\avg{n_{\alpha}'}\avg{y_{\alpha}}+Z$
can be solved with respect to $\alpha$, leading to $\kopt'$.
When the total number of possible variants $N$ is big,
$\avg{y_{\alpha}}\ll1$ and the term
$\avg{n_{\alpha}'}\avg{y_{\alpha}}$ can be neglected.
The approximate solution is then
\begin{equation}
\label{kopt-improved}
\kopt'\approx\frac{\ln(Z/Mp)}{\ln(1-p)}.
\end{equation}
This optimal number of variants to offer is smaller than $\kopt$
given by eq.~(\ref{noinfo-opt}). We can also notice that when
$p$ is small, using approximation $\ln(1-p)\approx-p$ we are
left with $\kopt'\approx\kopt$. This is an intriguing
property---by the two different approaches we obtained the same
result. To compare $\kopt'$ in markets with different sizes, we
plot it as a function of $Z/M$ in fig.~\ref{fig-improved}. As
can be seen, in a big market ($M\gtrsim100\,000$)
eq.~(\ref{kopt-improved}) fits well a numerical simulation of
the system.

One can examine also the increase of the vendor's profit caused
by the change of the sale method. Using previous results, the
approximate formula $\Delta\Xopt\approx Z[1+\ln(Mp/Z)]/(Np^2)$
can be obtained. We see that when the total number of variants
$N$ is big, sequential offering results in a small growth of the
vendor's profit. Nevertheless, in a system with a limited offer
(small $N$) or with very choosy buyers (very small $p$), the
improvement can be substantial.

Here we should notice, that the stopping condition ``income
greater than expenses'' introduced above can be hard to use in
practice. It is because $n_{\alpha}'$ is a random quantity and
can drop to the disadvantageous region
$n_{\alpha}'<Z+n_{\alpha}'y_{\alpha}$
even when $\avg{n_{\alpha}'}$ is big enough to cover the
expenses. Thus for the vendor it is not enough to simply check
profitability of the sale of one particular variant
$n_{\alpha}'$. Rather he has to take into account sales of all
previously offered variants. This is especially important in
systems with a small number of buyers $M$ where relative
fluctuations are bigger. This effect is shown in figure
\ref{fig-improved} where numerical results for the vendor
blindly using the stopping condition are shown for various
market sizes. Clearly as $M$ increases, numerical results
approach the analytical result (\ref{kopt-improved}).

\subsection{Competition of two vendors}
In real markets we seldom find a monopolist vendor; competition
and partition of the market is a natural phenomenon. To
investigate the model behavior in such a case we introduce the
second vendor to the market. We assume that the vendors differ
by initial costs, which are $Z_1$ and $Z_2$. Again we do not
consider the influence of advertisements and reputation, albeit
they are vital in a market competition.

\begin{figure}
\centering
\includegraphics[scale=0.26]{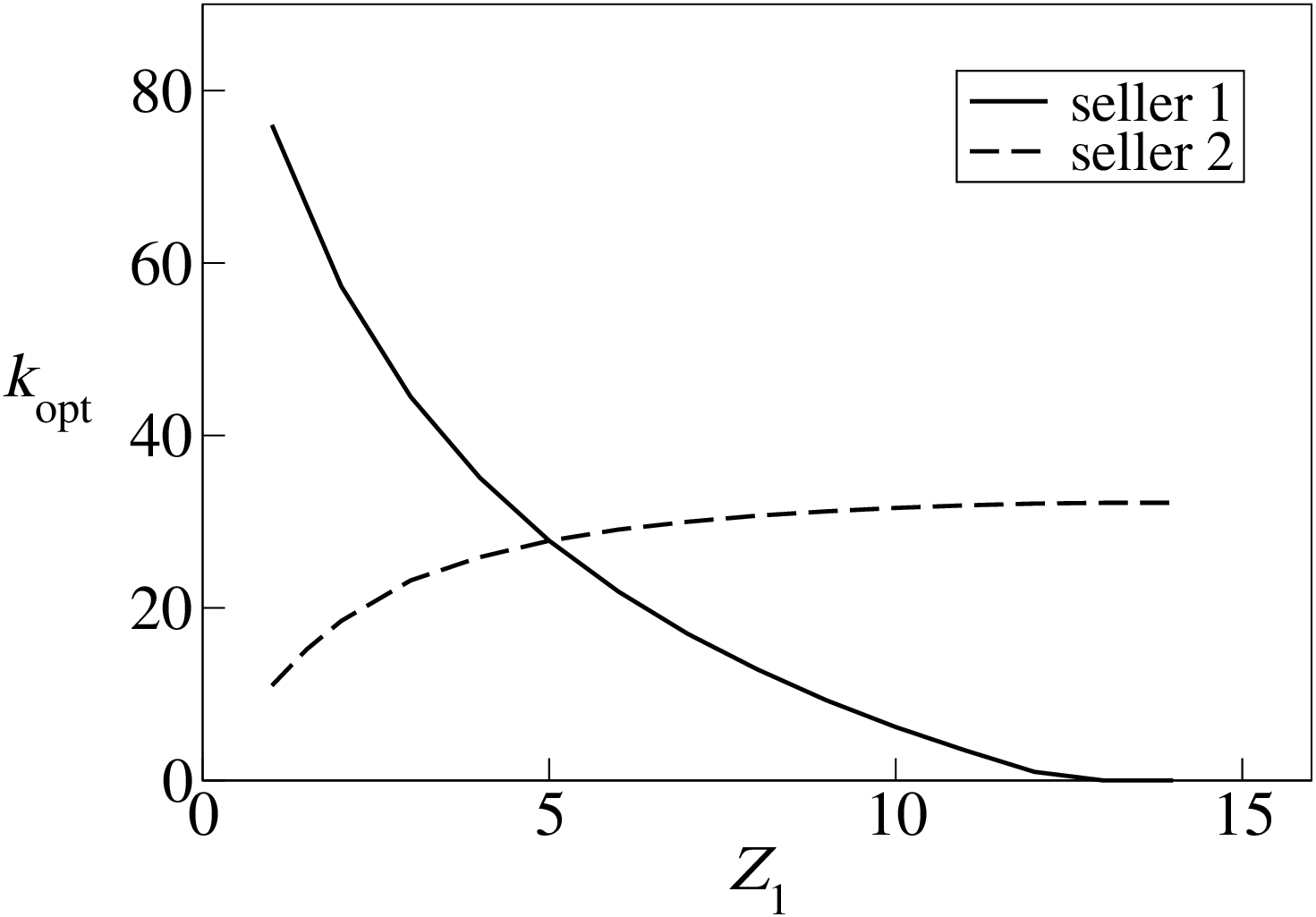}\quad
\includegraphics[scale=0.26]{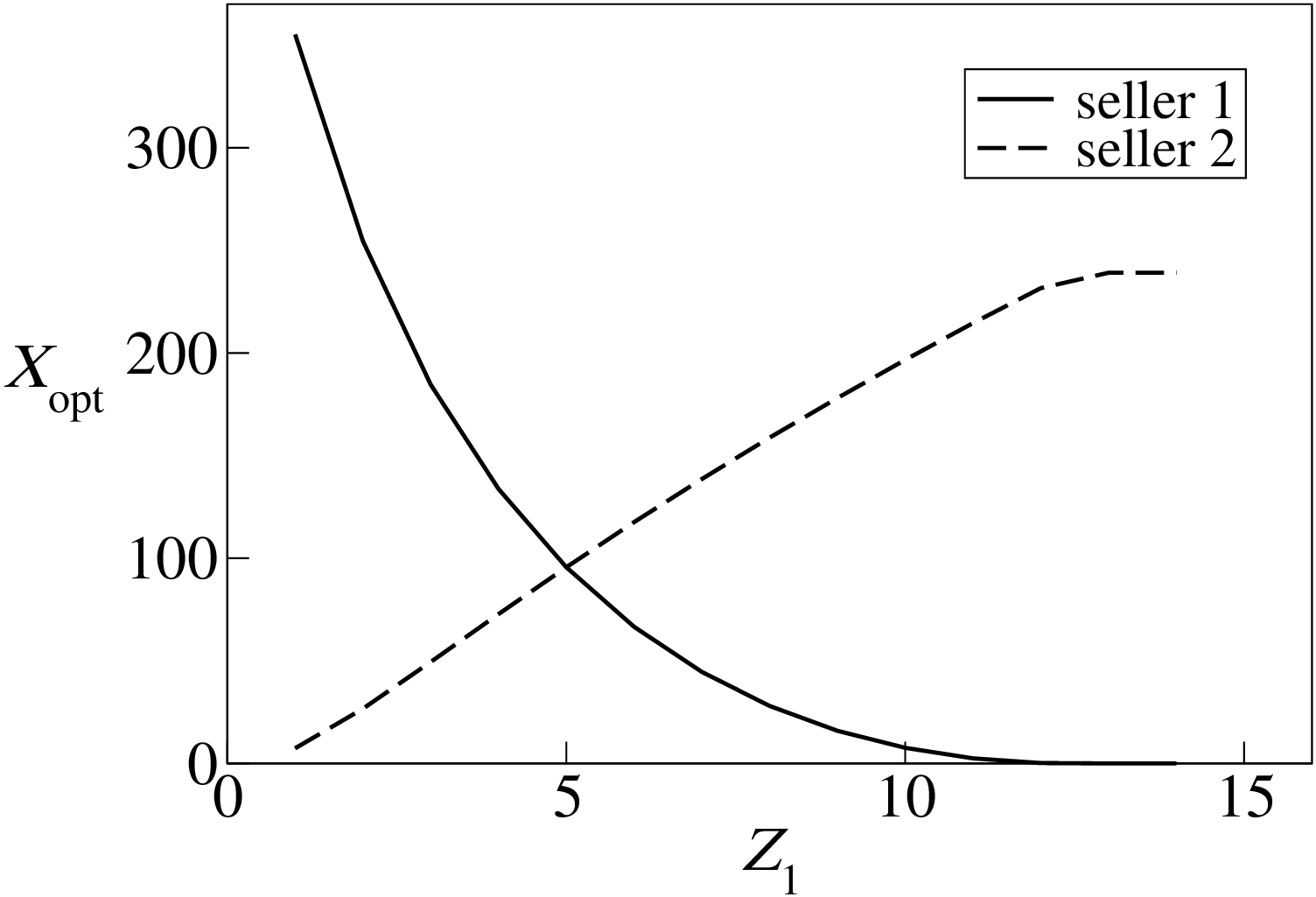}
\caption{The optimal number of variants to offer (left) and
the optimal profit (right) for vendor~1 (solid line) and for
vendor~2 (dashed line) against $Z_1$. The initial cost of the
second vendor is $Z_2=5.0$, $M=500$, and $p=0.05$.}
\label{fig-twovendors}
\end{figure}

The course of the solution is similar to the one leading to
eq.~(\ref{profit}). We label the number of variants offered by
vendor 1 as $k_1$, the number of variants offered by vendor 2 as
$k_2$, and we assume that there is no overlap between offered
variants. The aggregate sale of two buyers is $MP_A'$ where
\begin{equation*}
P_A'=1-(1-p)^{k_1+k_2}\approx 1-\exp[-p(k_1+k_2)].
\end{equation*}
With our assumptions about the equal status of the vendors,
every offered variant has the same average sale. Therefore both
vendors gain the share proportional to the number of variants
they offer. Thus vendor 1 takes $k_1/(k_1+k_2$ of the total
sale and vice versa. When $k_1/N,k_2/N\ll 1$, we can simplify
the expected profits to the form
\begin{subequations}
\label{twovendors}
\begin{align}
X_1(k_1,k_2)&=
M\big(1-\ee^{-p(k_1+k_2)}\big)\,\frac{k_1}{k1+k2}-k_1Z_1,\\
X_2(k_1,k_2)&=
M\big(1-\ee^{-p(k_1+k_2)}\big)\,\frac{k_2}{k1+k2}-k_2Z_2.
\end{align}
\end{subequations}
Both parties maximize their profits by adjusting $k_1$ and $k_2$.
The corresponding system
$\partial_{k_1}X_1(k_1,k_2)=0,\ \partial_{k_2}X_2(k_1,k_2)=0$
cannot be solved analytically but its numerical treatment is
straightforward. The result is shown in
fig.~\ref{fig-twovendors} where we have fixed the initial cost
$Z_2$ to investigate how $\kopt$ and $\Xopt$ for both vendors
vary with $Z_1$.

We see that at $Z_1\approx 12$ vendor 1 stops the production for
he cannot stand the competition of vendor 2. By putting $k_2=0$
in the equations $\partial_{k_1}X_1(k_1,k_2)=0$ and
$\partial_{k_2}X_2(k_1,k_2)=0$ we obtain the expression for the
value $Z_1^*$ when this price-out occurs
\begin{equation}
\label{two-vendors-stopping}
Z_1^*=\frac{Mp}{\ln(Mp/Z_2)}\bigg(1-\frac{Z_2}{Mp}\bigg).
\end{equation}
It is in a good agreement with the values found by a numerical
simulation of the model. Important feature of this result is
that it depends on the initial price $Z_2$ of the competitive
vendor---decreasing the production costs can expel others from
the market.

One can notice that when vendor 1 tries to increase the profit
by deliberately increasing $k_1$ (with the intention to increase
the sale), the term $-k_1Z_1$ prevents the success of this
strategy. As a result, the vendors have to adapt to each other.
In mathematic terms,
$X_1(k_{1\mathrm{opt}},k_{2\mathrm{opt}})\geq
X_1(k_1,k_{2\mathrm{ opt}})$.
At the same time, the sum of profits is not maximized at
$k_{1\mathrm{ opt}}$ and $k_{2\mathrm{ opt}}$. It is more
profitable to remove the less efficient producer (the one with
the higher value of initial costs). This is an analogy of a real
market where ruining (or taking over) of a competitor can
improve company profit.

\subsection{An informed vendor}
\label{sec-informed}
Now we would like to investigate the artificial case of the
market where the vendor knows costs $x_{i,\alpha}$ of all
buyers. This knowledge can be used to increase the optimal
profit. We start with a simpler question: if the vendor offers
only one variant, how much the sale can be increased by a good
choice of the variant?

The probability that buyer $i$ is agreeable to buy variant
$\alpha$ is $f(x_{i,\alpha})$. Since costs $x_{i,\alpha}$ are
random and independent, only the average acceptance probability
$p$ plays a role and the number of users willing to buy this
variant, $n_{\alpha}$, is thus binomially distributed with the
mean $\avg{n_{\alpha}}=Mp$ and the variance $\sigma^2=Mp(1-p)$.
When the number of buyers $M$ is big, we can pass to a
continuous approximation and assume the normal distribution of
$n_{\alpha}$
\begin{equation}
f(n_{\alpha})\approx\frac1{\sqrt{2\pi}\sigma}
\exp\left[-\frac{(n_{\alpha}-Mp)^2}{2\sigma^2}\right].
\end{equation}
The biggest value from the set $\{n_{\alpha}\}$
($\alpha=1,\dots,N$) we label as $m$. This is the number of
potential buyers for the most accepted variant and the vendor
does the best when by offering this variant. The probability
density $f_N(m)$ (often called \emph{extremal distribution}) is
\begin{equation}
\label{extremal-distrib}
f_N(m)=\frac{N}{\sqrt{2\pi}\sigma}\exp\bigg[
-\frac{(m-Mp)^2}{2\sigma^2}\bigg]\bigg(\frac12+
\frac12\,\mathrm{Erf}
\bigg[\frac{m-Mp}{\sigma\sqrt2}\bigg]\bigg)^{N-1}.
\end{equation}
The multiplication by $N$ appears because we do not care which
one of all $N$ variants is ``the most accepted'' one and the
error function term represents the probability that the
remaining $N-1$ variants are less accepted.

\begin{figure}
\centering
\includegraphics[scale=0.26]{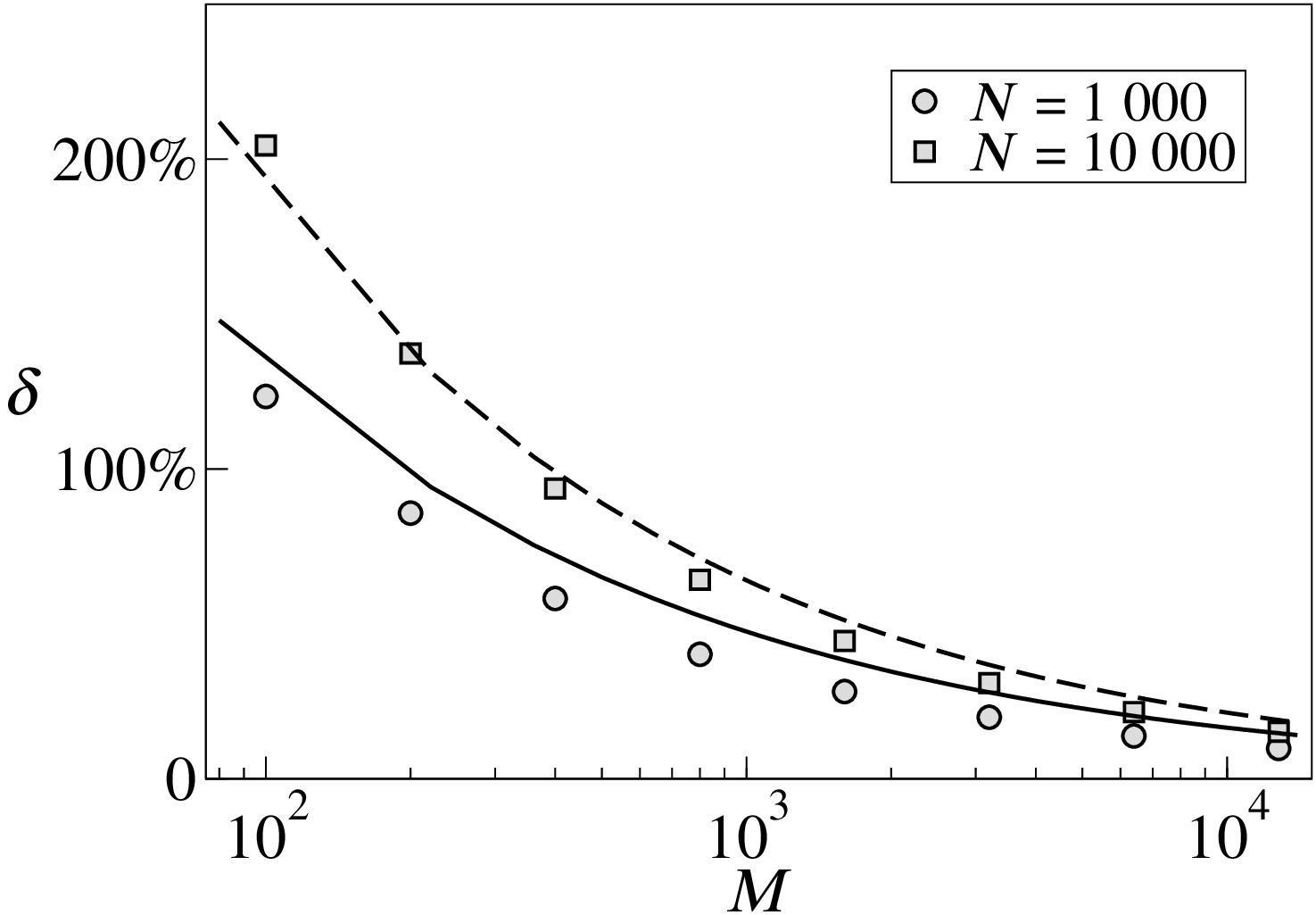}
\caption{The relative growth of the vendor's sale $\delta$ is
drawn against the total number of buyers $M$. Solid lines
represent the analytical result, outcomes from numerical
simulations are shown as symbols.}
\label{fig-informed}
\end{figure}

Since we are interested in big values of $N$, we expect that the
difference $\avg{m}-M$ is big in comparison with $\sigma$.
Therefore we use the approximation
$\mathrm{Erf}(x)\approx1-\exp[-x^2]/\sqrt{\pi x^2}$,
which is valid for $x\gg1$. When the error function value is
close to one, we can also use the approximation
$(1-x)^N\approx\exp[-Nx]$ ($x\ll1$) to obtain
\begin{equation*}
f_N(m)\approx\frac{N}{\sqrt{2\pi}\sigma}\exp\bigg[
-\frac{(m-Mp)^2}{2\sigma^2}-\frac{\sigma N}{\sqrt{2\pi}}
\frac{\exp\big[-(m-Mp)^2/2\sigma^2\big]}{m-Mp}\bigg].
\end{equation*}
This form is too complicated to obtain an analytical result for
$\avg{m}$. Instead we compute the most probable value $\tilde m$
\begin{equation*}
\tilde m\approx Mp+
\sigma\sqrt{2\ln\frac{\sigma^3N}{\sqrt{2\pi}}}\,.
\end{equation*}
Here the first term $Mp$ represents the average value of the
sale and the additional term represents the gain arising from
the additional vendor's knowledge. To get a better notion about
the sale growth we use the relative sale growth
\begin{equation}
\label{informed-growth}
\delta\equiv\frac{\tilde m-Mp}{Mp}\approx
\sqrt{\frac1{Mp}\ln\frac{pMN^2}{2\pi}}.
\end{equation}
To simplify the formula, the assumption $p\ll1$ has been used.
A comparison of this result with a numerical simulation of the
model is shown in fig.~\ref{fig-informed}. As can be seen, a
good agreement is obtained.

When $\delta\ll1$, all vendor's information is indeed useless
and the average sale improvement is negligible. The inequality
$\delta\ll1$ leads to the condition
\begin{equation}
\label{useless-info}
N^2\ll\frac{2\pi}{Mp}\,\ee^{Mp}.
\end{equation}
Thus when the number of variants is not large enough, buyers'
opinions in the uncorrelated market cannot be used to increase
the vendor's sale and profit.

From the previous results we can draw useful implications about
the vendor with perfect information, offering more than only one
variant. When the total number of variants $N$ is big, the
number of variants offered by the vendor is small in comparison
with $N$. Therefore the average sale of all offered variants is
increased at most by $\delta$ given by
eq.~(\ref{informed-growth}) and the same applies to the total
sale. However, the vendor is interested mainly in his
profit. When we take into account different costs $y_{\alpha}$
of variants, the resulting growth of the income due to the
informations is even smaller than $\delta$ because the variant
with the highest sale can have a high cost for the vendor. Thus
condition (\ref{useless-info}) is has more general consequences.
It specifies the circumstances when even the perfect information
about buyers' preferences do not help the vendor to achieve a
significant improvement of his profit.

\section{Correlations in the system}
Now we would like to add one important flavor to the
model---correlations. They arise from conformity of people's
tastes (buyer-buyer correlations) and from the fact that high
quality preferred by buyers results in high costs on the
vendor's side (buyer-vendor anticorrelations). To approach the
behavior of a real market, we investigate how these correlations
influence our results obtained so far. Before doing so, we
briefly discuss correlations from a general point of view.

\subsection{Measures of correlations}
A correlation is the degree to which two or more quantities are
associated. We shall discuss different ways how to measure
correlations and how to introduce them to the system. In
particular, we would like to measure the correlation between two
lists (vectors) of costs: $\vec{x}_i$ and $\vec{x}_j$ (two
buyers) or $\vec{x}_i$ and $\vec{y}$ (a buyer and the vendor).
All lists of our interest have length $N$ and contain real
numbers between $0$ and $1$. A common choice for the correlation
measure is Pearson's correlation coefficient $r$. For lists
$\vec{x}$ and $\vec{y}$ it is defined as
\begin{equation}
\label{pearson}
r^2=\frac{\Big[\sum_{\alpha=1}^N
(x_{\alpha}-\overline{x})(y_{\alpha}-\overline{y})\Big]^2}
{\sum_{\alpha=1}^N (x_{\alpha}-\overline{x})^2
\sum_{\alpha=1}^N (y_{\alpha}-\overline{y})^2}.
\end{equation}
This measure is sensitive to non-linear transformations of
values in lists $\vec{x}$ and $\vec{y}$. In addition, since it
originates in the least-square fitting of the data by a straight
line, it measures only a linear correlation. For this reasons, in
this work we use another correlation measure, Kendall's tau. For
lists $\vec{x}$ and $\vec{y}$ it is given by the formula
\begin{equation}
\label{kandall}
\tau=\frac2{N(N-1)}\sum_{\alpha<\beta}
\sigma_{\alpha\beta},\qquad
\sigma_{\alpha\beta}=
\mathrm{sgn}\,[(x_{\alpha}-x_{\beta})(y_{\alpha}-y_{\beta})]
\end{equation}
and it ranges from $+1$ (exactly the same ordering of lists
$\vec{x}$ and $\vec{y}$) to $-1$ (reverse ordering of lists);
uncorrelated lists have $\tau=0$. Notably, Kendall's tau is
insensitive to all monotonic mappings of the data. This is the
strongest property we can expect from a correlation
measure---more general transformations, nonmonotonic mappings,
can sweep out any structure present in the data.

\subsection{Lists with a given correlation degree}
\label{sec-giventau}
Now we would like to construct a set of lists that have mutual
values of Kendall's tau equal to $\tau_0$. Such a set would
represent lists of buyers' preferences in an equally dispersed
society. Since buyers' tastes are to a certain extent similar,
we expect positive correlations with $\tau_0>0$. Nevertheless,
in the following discussion we do not confine ourself to this
region.

First we address a different question. Let's assume that between
lists $1$ and $2$ there is $\tau_{12}$, between lists $1$ and
$3$ there is $\tau_{13}$. Does it imply any constraints on
$\tau_{23}$? The answer is yes. It can be shown (see Appendix
\ref{appendix-proof}) that $\tau_{23}$ fulfills the inequality
\begin{equation}
\label{tauinequality}
\abs{\tau_{12}+\tau_{13}}-1\leq\tau_{23}\leq
1-\abs{\tau_{12}-\tau_{13}},
\end{equation}
which is an analogy of the triangular inequality for side
lengths of a triangle. From (\ref{tauinequality}) we can draw
various simple conclusions. First, if we want to construct three
lists which have pairwisely $\tau_0$, it is possible only for
$-1/3\leq\tau_0\leq1$.\footnote{An example for lists with the
pairwise value $\tau_0=-1/3$: $x_1=\{3,2,1\}$, $x_2=\{2,1,3\}$
and $x_3=\{1,3,2\}$.}
Thus it is impossible to have more than two lists which are
perfectly anticorrelated. Another simple result is that when
$\tau_{12}=-1$, inevitably $\tau_{23}=-\tau_{13}$.

\begin{figure}
\centering
\includegraphics[scale=0.26]{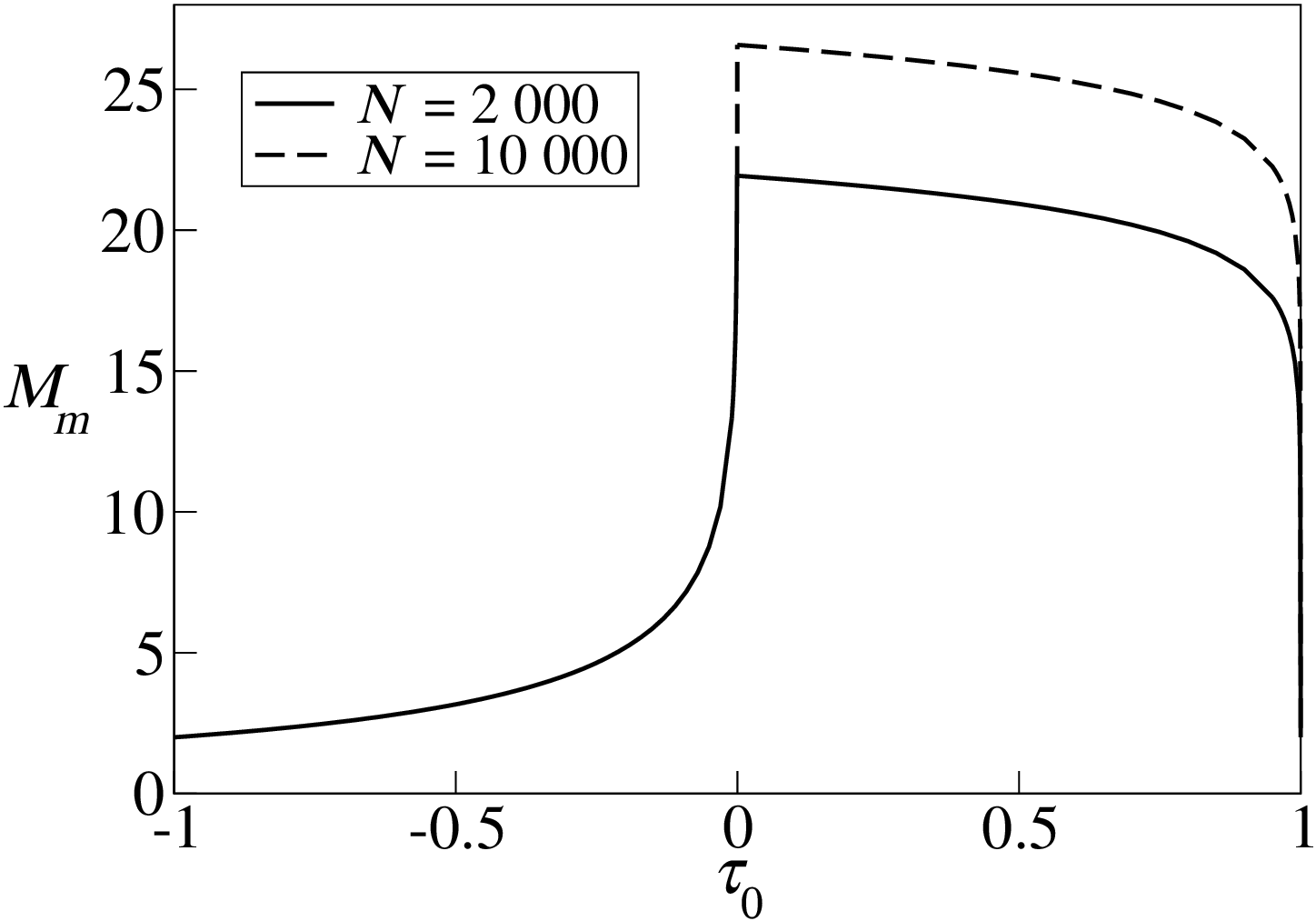}
\caption{The upper bound $M_m$ as a function of $\tau_0$ for
two different lengths of lists $N$. In both cases the upper
bound $M_m$ is the same over a large part of the region
$\tau_0<0$ and drops to $1$ when $\tau_0\to1$.}
\label{fig-M_m}
\end{figure}

Now the question is whether we are able to create the whole
system of $M$ lists which all have pairwisely Kendall's tau
equal to $\tau_0$. The answer depends on the magnitude of $M$.
It can be shown (see appendix \ref{appendix-ubound}) that the
upper bound for $M$ is
\begin{equation}
\label{max-size}
M_m=\begin{cases}
2+\log_2\frac{(1-\tau_0)(N-1)N}4 &
(\tau_0\geq0),\\
\min\big[2+\log_2\frac{(1-\tau_0)(N-1)N}4,\ 
2\log_2\frac{1-\tau_0}{-\tau_0}\big] & (\tau_0<0).
\end{cases}
\end{equation}
As can be seen in fig.~\ref{fig-M_m}, this quantity grows slowly
with the list length $N$. Therefore to model a market with a
large number of equally correlated buyers we would need an
enormous number of possible variants.

\subsection{Generation of correlated lists}
In the previous paragraphs we found that the society with
a fixed mutual correlation degree of buyers is limited in its
size. Therefore to introduce correlations to the presented
market model we need a different approach. While \emph{copulas}
represent a general tool (see e.g.~\cite{copulas,copulas2}),
they are useful mainly for numerical simulations and offer only
small possibilities for analytical results. Here we adopt a
simpler way to generate correlated lists.

Let's consider the lists of variables
\begin{subequations}
\label{lists}
\begin{align}
x_{i,\alpha}&=(1-t)\,a_{i,\alpha}+t\,c_{\alpha},\\
y_{\alpha}&=(1-t)\,b_{\alpha}+st\,c_{\alpha}+\frac t2(1-s),
\end{align}
\end{subequations}
where $a_{i,\alpha}$, $b_{\alpha}$ and $c_{\alpha}$ are
independent random variables uniformly distributed in the range
$[0;1]$. Here $s=\pm1$ indicates correlation/anticorrelation
between $\vec{x}_i$ and $\vec{y}$ and $t\in[0;1]$ is the binding
parameter controlling strength of the correlation: $t=0$ leads
to uncorrelated lists, $t=1$ to perfectly correlated ($s=1$) or
anticorrelated ($s=-1$) lists. In all cases, values
$x_{i,\alpha}$, $y_{\alpha}$ lie in the range $[0;1]$.

\begin{figure}
\centering
\includegraphics[scale=0.26]{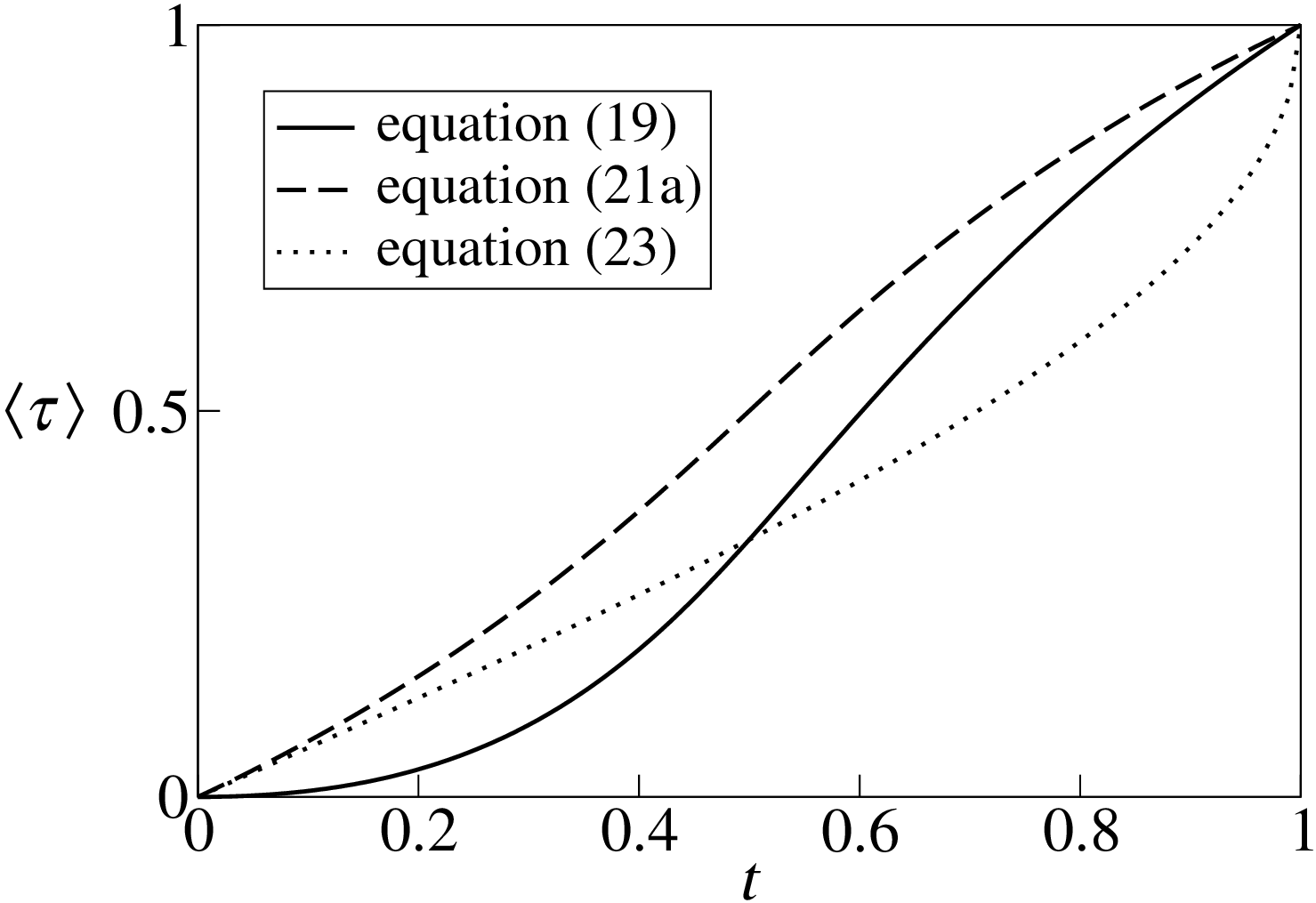}
\caption{The dependence of $\avg{\tau}$ on $t$ for the proposed
constructions of correlated lists.}
\label{fig-tau}
\end{figure}

For the lists defined above, it can be shown that (see appendix
\ref{appendix-tau})
\begin{equation}
\label{avgtau}
\avg{\tau_{xy}}=s\,\avg{\tau_{xx}},\quad
\avg{\tau_{xx}}=\begin{cases}
\frac{u^2}{15}\,\big(10-6u+u^2\big) & (u\leq1),\\
\frac1{15}\,\big(15-\tfrac{14}u+\tfrac4{u^2}\big) & (u>1),
\end{cases}
\end{equation}
where $t/(1-t)\equiv u$. Plots of $\avg{\tau_{xx}}$ and
$\avg{\tau_{xy}}$ are shown in fig.~\ref{fig-tau}. Since buyers'
lists are prepared using the same formula, the average value of
their correlation is non-negative. Notably, for any value of
$\tau$ we can find suitable $s$ and $t$ that produce lists with
the expected correlation equal to $\tau$.

Lists created using eq.~(\ref{lists}) do not have fixed mutual
correlation, its actual value fluctuates around the mean value
given by (\ref{avgtau}). According to the law of large numbers,
$f(\tau)$ is normally distributed. In
appendix~\ref{appendix-tau} it is shown that the variance of
$\tau$ is proportional to $1/N$. Such fluctuations are
negligible for long lists. We can conclude eq.~(\ref{lists})
present a way to create a system with the desired amount of
correlation $\tau$ for any $\tau$, $M$ and $N\gg1$.

Yet there is a hitch in the proposed construction of correlated
lists. The parameter $t$ influences the distribution of costs:
for $t=0$ or $t=1$ they are distributed uniformly, for $t=0.5$
$f(x)$ has a tent shape. This is an implausible property: the
changes of the cost distributions can drive or distract the
phenomena we are interested in. To fix this problem we propose
the following two solutions.

First, to obtain correlated lists we can use the formulae
\begin{equation}
\label{lists2}
x_{i,\alpha}=
\frac12+st\,\bigg(\frac{\alpha-1}{N-1}-\frac12\bigg)+
(1-t)\,\bigg(s_{i,\alpha}-\frac12\bigg),\quad
y_{\alpha}=\frac{\alpha-1}{N-1},
\end{equation}
where $s_{\alpha,j}$ is a random quantity distributed uniformly
in the range $[0;1]$. The complicated form of the $x_{i,\alpha}$
has a simple meaning. The vendor's costs grow uniformly with
$\alpha$ and buyers' costs are connected to the vendor's by the
parameter $t\in[0;1]$. The term proportional to $1-t$
introduces a noise to the system, resulting in differences
between buyers' and vendor's lists. Finally, the term $1/2$
represents the average value of buyers' costs. It is easy to
check that $x_{i,\alpha}$ given by (\ref{lists2}) is confined to
the range $[0;1]$ for every $t\in[0;1]$ and $s=\pm1$. The
overall distribution of costs is uniform in the range $[0;1]$
and thus we avoid the problems of eq.~(\ref{lists}). Moreover,
this construction is simple enough to tract the proposed model
analytically.

Using the techniques shown in appendix \ref{appendix-tau} we can
find Kendall's tau in this case. In the limit $N\to\infty$ one
obtains
\begin{subequations}
\label{avgtau4}
\begin{align}
\avg{\tau_{xy}}&=\begin{cases}
\frac s6\big(4u-u^2\big) & (u\leq1),\\
\frac s6\big(6-\tfrac4u+\tfrac1{u^2}\big) & (u>1),
\end{cases}\\
\avg{\tau_{xx}}&=\begin{cases}
\frac{u^2}{15}\big(10-6u+u^2\big) & (u\leq1),\\
\frac1{15}\big(15-\tfrac{14}u+\tfrac4{u^2}\big) & (u>1),
\end{cases}
\end{align}
\end{subequations}
where again $u\equiv t/(1-t)$. The form of $\avg{\tau_{xx}}$ is
identical with (\ref{avgtau}) found before for a different
construction of correlated lists.

As we will see later, eq.~(\ref{lists2}) is not appropriate to
produce anticorrelated lists. Hence we present one more approach
here---less accessible to analytical computation but more
robust. The normal distribution is stable with respect
to addition of random variables and this motivates us to make
the following choice
\begin{subequations}
\label{lists-gauss}
\begin{align}
x_{i,\alpha}&=\sqrt{1-t}\,a_{i,\alpha}+\sqrt{t}\,c_{\alpha},\\
y_{\alpha}&=\sqrt{1-t}\,b_{\alpha}+s\sqrt{t}\,c_{\alpha}
\end{align}
\end{subequations}
where $a_{i,\alpha},b_{\alpha},c_{\alpha}$ are drawn from the
standard normal distribution. It can be shown\footnote{In the
derivation, the following formula is useful
$$
\int_0^{\infty}\ee^{-px^2}
\mathrm{Erf}(ax)\,\mathrm{Erf}(bx)\,\dd x=
\frac1{\sqrt{\pi p}}\arctan\Big[ab/\sqrt{p(a^2+b^2+p)}\,\Big].
$$}
that in this case
\begin{equation}
\label{avgtau3}
\avg{\tau_{xx}}=\frac2{\pi}\arcsin t,\quad
\avg{\tau_{xy}}=s\,\avg{\tau_{xx}}.
\end{equation}
The course of $\avg{\tau_{xx}}$ is shown in fig.~\ref{fig-tau}.

As our market model assumes costs confined to the range $[0;1]$,
the costs given by (\ref{lists-gauss}) have to be transformed
using the cumulative distribution function of the standard
normal distribution $\Phi(x)$. In this way we obtain
\begin{equation}
\label{transformation}
\hat x_{i,\alpha}=\Phi^{-1}(x_{i,\alpha}),\quad
\hat y_{\alpha}=\Phi^{-1}(y_{\alpha}).
\end{equation}
Since this transformation is monotonic, it does not affect the
value of $\avg{\tau}$ and we can use eq.~(\ref{avgtau3}) for
transformed lists of costs.

\section{A market with correlations}
When we discussed the market without correlations, the
probability distribution of the variant cost
$\pi_{\alpha}(x_{i,\alpha})$ was independent of $\alpha$.
Consequently, the probability of accepting variant $\alpha$
\begin{equation}
\label{acceptance-prob}
P_A(\alpha)=
\int_D\pi_{\alpha}(x_{i,\alpha})f(x_{i,\alpha})\dd x_{i,\alpha}
\end{equation}
was also independent of $\alpha$ (we labeled $P_A\equiv p$). As
a result, when we change the acceptance function $f(x)$ while
preserving the quantity $\int_0^1 \pi(x)f(x)\dd x$, the derived
results remain unchanged.

In the presence of correlations we witness a very different
picture: the detailed shape of the acceptance function $f(x)$ is
important. To keep the algebra as simple as possible, from now
on we adopt the simplest choice for $f(x)$: the step function
$f(x)=1-\varTheta(x-p)$. This means that a buyer accepts a
proposed variant only when its cost is smaller than $p$.

In the following we first deal with the market where costs are
given by eq.~(\ref{lists2}) for it is more accessible to
analytical treatment. Then we shortly present analytical results
for the market where to introduce correlations,
eq.~(\ref{lists-gauss}) is used.

\subsection{An uninformed vendor in a market with correlations}
Here we assume cost correlations created using
eq.~(\ref{lists2}). When the vendor has no information about the
preferences of buyers, similarly to sec.~\ref{sec-noinfo} the
best strategy is to produce vendor's most favorable variants.
First we focus on the case of positive correlations; in
(\ref{lists2}) we set $s=1$ and $0\leq t\leq 1$. Using
(\ref{acceptance-prob}) and the chosen step acceptance function
$f(x)$, the probability that one buyer accepts variant $\alpha$
is
\begin{equation}
\label{PA2}
P_A(\alpha)=
\begin{cases}
0 & 1+(N-1)\,\frac pt<\alpha,\\
\frac1{1-t}\big[p-t\frac{\alpha-1}{N-1}\big] &
1+(N-1)\frac{p+t-1}t<\alpha<1+(N-1)\,\frac pt,\\
1 & \alpha<1+(N-1)\frac{p+t-1}t.
\end{cases}
\end{equation}
Since we expect the total number of variants $N$ to be very
large and $p$ rather small, the second region makes the major
contribution and thus we simplify eq.~(\ref{PA2}) to
$P_A(\alpha)\approx(p-t\alpha/n)/(1-t)$.

We assume that the vendor is simultaneously offering his $k$
most favorable variants. The probability that one buyer denies
all offered variants is
\begin{equation}
P_D(k)=\prod_{\alpha=1}^k\big[1-P_A(\alpha)\big]=
\prod_{\alpha=1}^k\bigg(1-\frac{p}{1-t}\bigg)
\bigg(1+\frac{t\alpha}{N(1-t+p)}\bigg).
\end{equation}
Since $N$ is big, we use the approximation $1-x\approx\exp[-x]$
to evaluate this expression analytically, leading to
\begin{eqnarray}
P_D(k)&\approx&\bigg(1-\frac{p}{1-t}\bigg)^k\prod_{\alpha=1}^k
\exp\bigg[-\frac{t\alpha}{N(1-t+p)}\bigg]\approx\nonumber\\
&\approx&\exp\bigg[-\frac{pk}{1-t}+\frac{tk^2}{2N(1-t+p)}\bigg].
\label{P_D}
\end{eqnarray}
Here we used also $1-p/(1-t)\approx\exp[-p/(1-t)]$ which is
valid when $p/(1-t)$ is small. When this is not the case,
denying probability $P_D(k)$ approaches zero and thus accepting
probability is virtually one regardless to the approximation
used.

With respect to (\ref{lists2}), the sum of expected vendor's
costs can be written as
\begin{equation}
\label{costs}
kZ+\sum_{\alpha=1}^k MP_S(\alpha)\frac{\alpha-1}{N-1}\approx
kZ+\sum_{\alpha=1}^k MP_S(\alpha)\frac{\alpha}N.
\end{equation}
Here the first term represents fixed costs for producing $k$
different variants, $P_S(\alpha)$ is the probability that
to one buyer variant $\alpha$ is sold. Since the probability
that of the successful trade is $1-P_D(k)$, from the condition
$\sum_{\alpha=1}^k P_S(\alpha)=1-P_D(k)$ we can deduce
\begin{equation}
\label{P_S}
P_S(\alpha)=\frac{P_A(\alpha)}{\sum_{\alpha=1}^k P_A(\alpha)}\,
\big[1-P_D(k)\big].
\end{equation}
This corresponds to the portioning of the probability $1-P_D(k)$
among $k$ variants according to their probability of acceptance.

\begin{figure}
\centering
\includegraphics[scale=0.26]{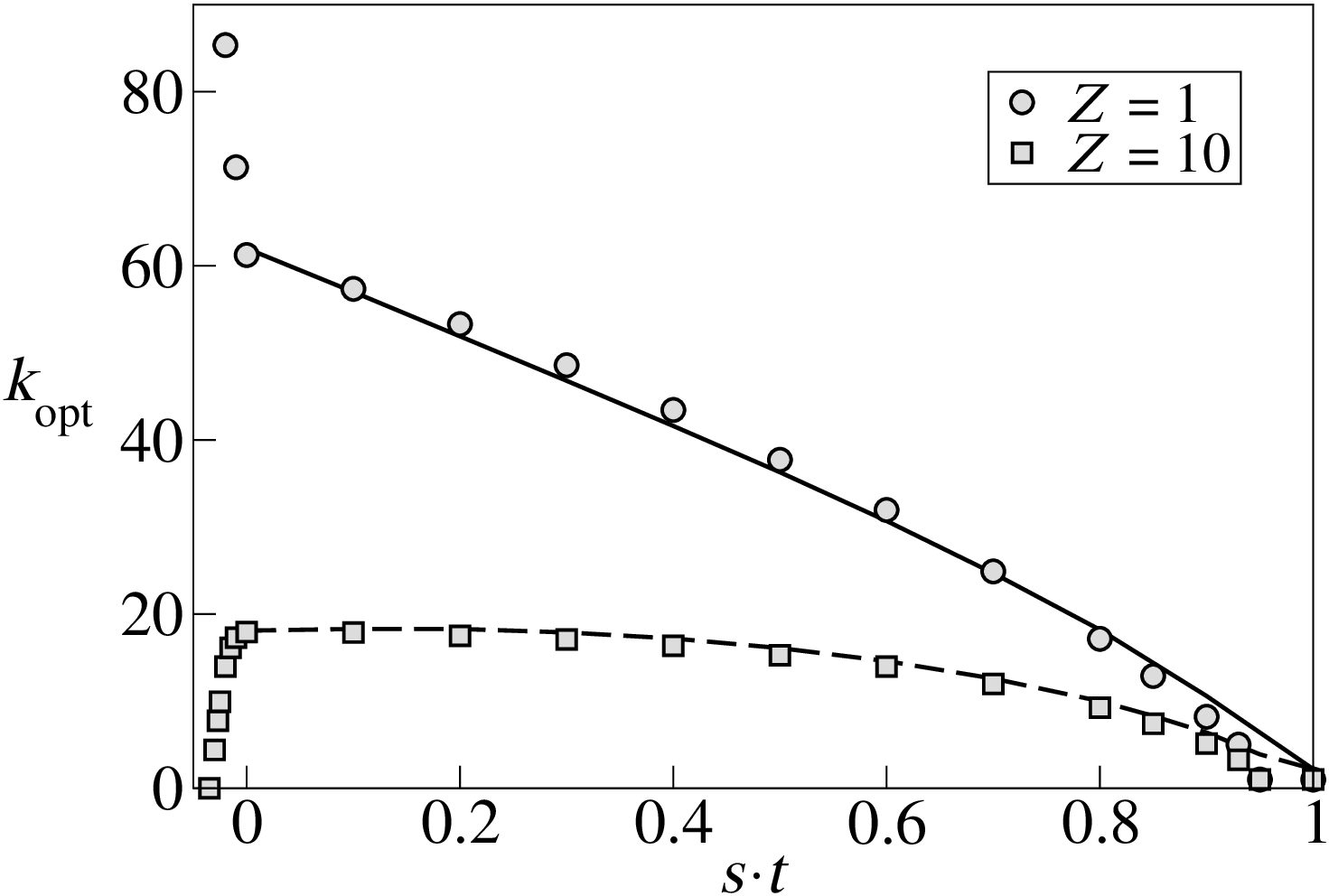}\quad
\includegraphics[scale=0.26]{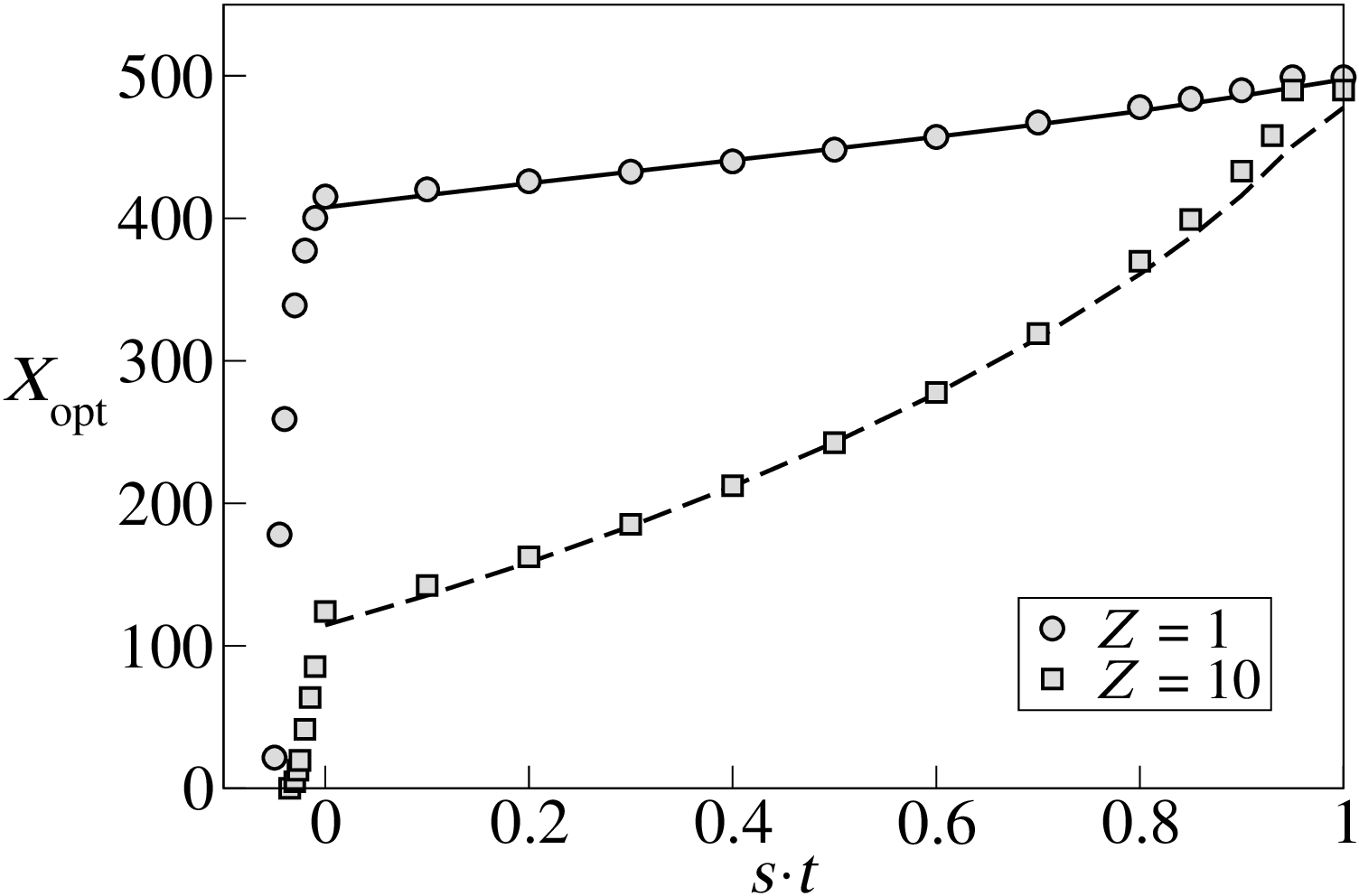}
\caption{The optimal number of variants to produce (left) and
the optimal profit drawn against $st$ for two different values
of the initial cost $Z$. Lines show analytical results derived
above, symbols represent numerical simulations (averages of
1\,000 realizations), model parameters are set to $N=2\,000$,
$M=500$, $p=0.05$. The decay of both quantities for $st<0$ is in
agreement with eq.~(\ref{threshold}).}
\label{fig-correlated}
\end{figure}

Now we can use (\ref{P_D}), (\ref{costs}) and (\ref{P_S}) to
write down the expected profit of the vendor offering his $k$
topmost variants $\overline{X}(k)$. It's not possible to carry
out the maximization of this expression analytically---numerical
techniques have to be used to find $\kopt$ and $\Xopt$. Results
are shown in \ref{fig-correlated} as lines together with
outcomes from a numerical simulation of the model; a good
agreement is found for $st>0$. Results confirm that positive
correlations between buyers and the vendor increase the vendor's
profit. This pattern is most obvious in the case $t=1$ when the
vendor can offer only the most favorable variant and still every
buyer buys it.

In the numerical results shown in fig.~\ref{fig-correlated} we
can notice one striking feature. When $st<0$, $\kopt$ changes
rapidly and $\Xopt$ falls to zero quickly. Such a behavior is
rather surprising for one do not expect abrupt changes in the
region $st<0$ when there were none in the opposite region
$st>0$. The reason for this behavior is simple---when $s=-1$,
vendor's most preferred variants have cost too high to be
accepted by a buyer. This effect can be quantified. When buyers'
costs are generated by (\ref{lists2}), the inequality
$x_{i,\alpha}\geq t(N-\alpha)/(N-1)$ holds. Due to the
acceptance function only the variants with cost smaller than $p$
are accepted. Therefore only variants with $\alpha\geq\amin$ can
be possibly accepted, where
\begin{equation}
\label{threshold}
\amin=1+(N-1)\,\frac{t-p}{t}\approx N(1-p/t).
\end{equation}
Thus with negative correlations in the market, the vendor is
able to sell the most favorable variant (the one with
$\alpha=1$) only if $p\geq t$. When $p<t$, the vendor sells no
variants $\alpha=1,\dots,\amin-1$. Since $\amin$ grows steeply
with $t$ (already with $t=2p$ one obtain $\amin=N/2$), the
vendor offering his top $k$ variants has to offer too many of
them and he suffers both big initial costs and big costs
$y_{\alpha}$. As a result the vendor is pushed out of the
market.

Without detailed investigation we can infer the system behavior
when the step acceptance function is replaced by a different
choice. In the limit of careless buyers with $f(x)=C$, the
influence of correlations vanishes and both $\kopt$ and $\Xopt$
do not depend on $st$ and the model simplifies to the case
investigated in sec.~\ref{sec-noinfo}. Thus as $f(x)$ gradually
changes from the step function to a constant function, the
dependence on $st$ gets weaker. In particular, if the largest
cost $x$ for which $f(x)>0$ is $x_0$ (for the step function
$x_0=p$), in eq.~(\ref{threshold}) $p$ is replaced by $x_0$. As
a consequence, $\amin=$ decreases and the steep decline of
$\Xopt$ in fig.~\ref{fig-correlated} shifts to a lower value of
$st$.

\subsection{An uninformed vendor in a different market with
correlations}
\begin{figure}
\centering
\includegraphics[scale=0.26]{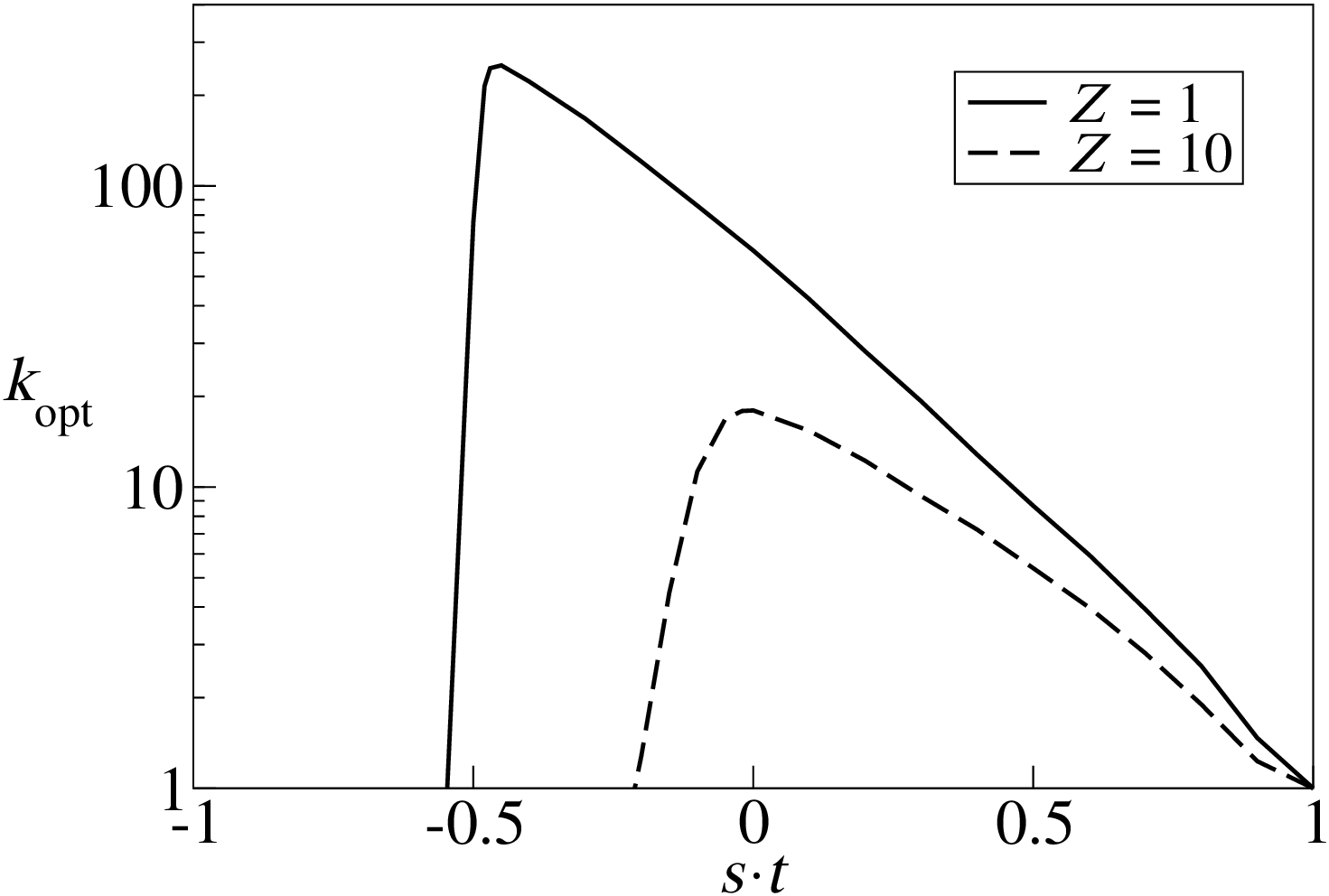}\quad
\includegraphics[scale=0.26]{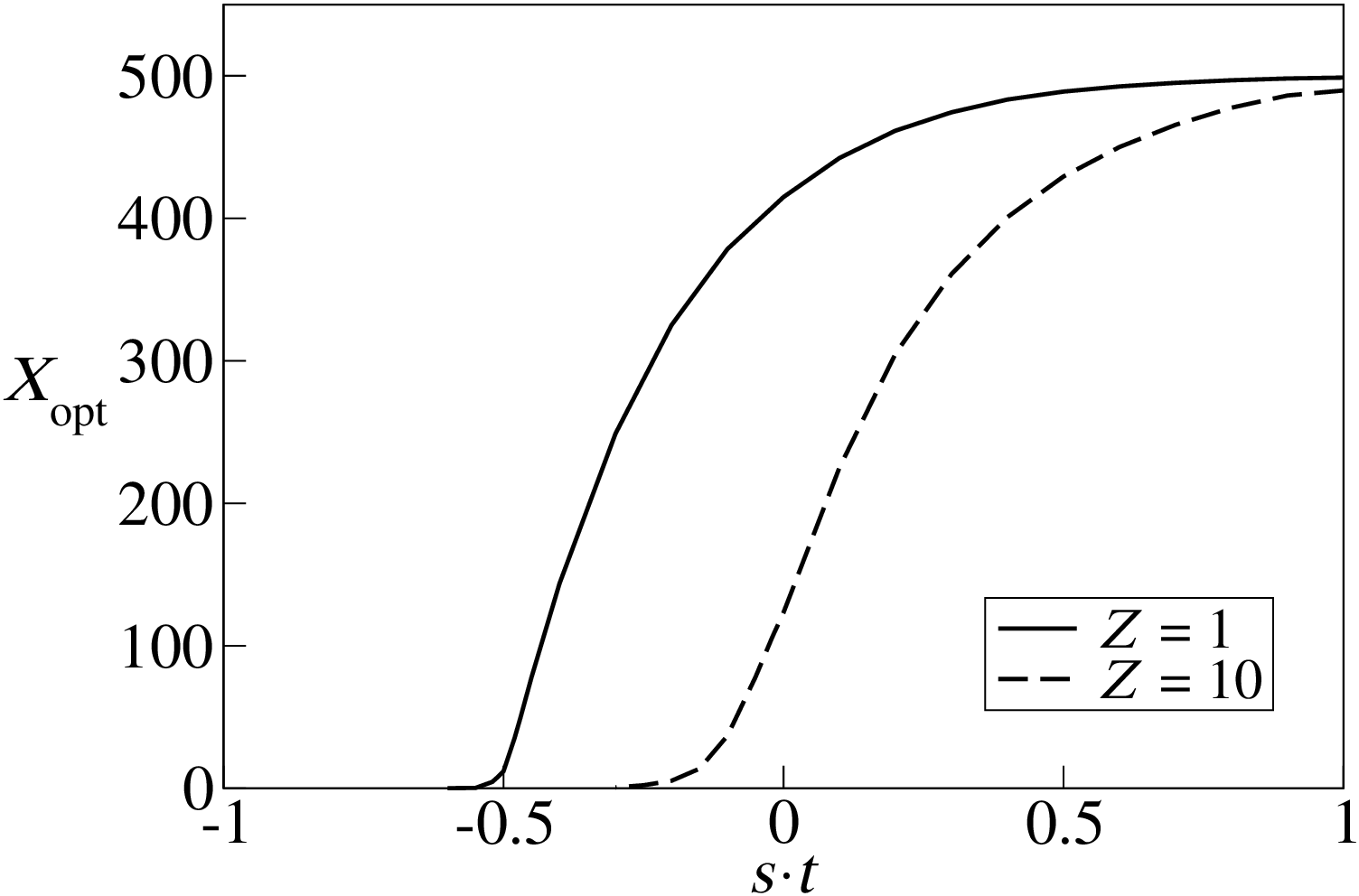}
\caption{The optimal number of variants and the optimal profit
of the uninformed vendor in the market with correlations given
by (\ref{lists-gauss}) and (\ref{transformation}). Numerical
results are averages of 1\,000 repetitions, $p=0.05$,
$N=2\,000$, $M=500$.}
\label{fig-correlated-gaussian}
\end{figure}

Now we switch to the market costs drawn using
eq.~(\ref{lists-gauss}) and transformed to the range $[0;1]$ by
eq.~(\ref{transformation}). As we already mentioned, this case
is not allowable for an analytical treatment---hence we present
only numerical results in fig.~\ref{fig-correlated-gaussian}.
They agree with our expectations: when $st=1$, for the vendor
it's sufficient to produce only one variant. As the positive
correlations diminish, the optimal number of offered variants
grows and the profit shrinks. A closer investigation of the
vendor's behavior in this case exceeds the scope of this paper
and remains as a future challenge.

\section{Another trading model}
\label{sec-model2}
In previous sections we presented a way how to deal with the
trading process. Here we shortly present a different model which
arises from the same playground as our previous reasonings but
highlight slightly different aspects of the market phenomenon.

Let's have a market with $M$ buyers and $N$ different variants
that the vendor can produce. Preferences of the interested
parties are again represented by the scalar costs
$x_{i,\alpha}$, $y_{\alpha}$ ($i=1,\dots,M$, $\alpha=1,\dots,N$)
uniformly distributed in the range $[0;1]$ (thus again we have
no correlations in the system).

In a market, a vendor is aware that when some buyer is not
satisfied with the offer, she can choose a different vendor.
Therefore every vendor tries to induce as small cost as possible
to the customers. This can be done by offering of the variants
highly preferred by many buyers. We can visualize the process by
sorting preference lists of all interested parties. Now when
some variant is near the top of a buyer's list, its cost is
small and it is favorable for this buyer.

\begin{figure}
\centering
\includegraphics[scale=1]{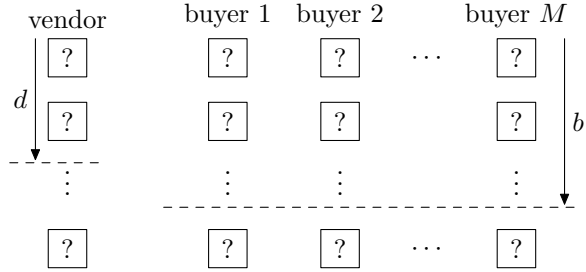}
\caption{The layout of the introduced trading model. Each column
represents a sorted list of variants' costs (the most preferred
at top). The vendor is willing to go down his list by $d$, in
consequence buyers are forced go down by some value $b$.
Questionmarks signalize that after sorting of all lists we do
not know standing of variants in the lists.}
\label{fig-layout}
\end{figure}

We have to specify the criterion for the ``variant preferred by
many buyers''. First, it can be the variant that is not too deep
in nobody's list. Thus, if we label the position of variant
$\alpha$ in the list of buyer $i$ as $k_{i,\alpha}$ and
$\max_i k_{i,\alpha}$ as $b_{\alpha}$, the vendor chooses the
variant $\alpha$ that has the smallest $b_{\alpha}$. The
selection process is visualized in fig.~\ref{fig-layout}. Now
the question is: how far buyers have to go down their lists? In
other words: if we label $b\equiv\min_{\alpha}b_{\alpha}$, what
is $\avg{b}$?

Since we have $M$ buyers in the market, the probability of
a~particular value $b$ is approximately given by the formula
\begin{equation}
\label{walk-prelim}
P(b)\approx\bigg[1-\bigg(\frac{b-1}{N}\bigg)^M\bigg]^d\times
\bigg[M\frac dN\bigg(\frac bN\bigg)^{M-1}\bigg].
\end{equation}
Here the first term denotes the probability that there is no
such a~variant which is among topmost $b-1$ for every buyer and
among topmost $d$ for the vendor.\footnote{Since in one list
each variant appears only once, this is only an approximate form
of the probability.}
The second term responds to the fact that there is some variant
which is exactly on $b$th place in the list of a buyer (we do
not care who it is, thus the multiplication by $M$ appears),
among $b$ topmost variants in lists of other buyers and among
$d$ topmost variants in the vendor's list.

In eq.~(\ref{walk-prelim}) we can use the approximation
$(1-x)^s\approx\exp[-xd]$ which is valid when $x\ll1$. To
calculate $\avg{b}=\sum_{b=1}^N bP(b)$ we replace the summation
by the integration in the range $[0;\infty]$ which yields
\begin{equation}
\label{walk-b}
\avg{b}\approx\frac{N\Gamma(1/M)}M\,d^{-1/M}.
\end{equation}
Here we dropped terms vanishing in the limit $N\to\infty$ (for
there is a big number of variants that the vendor can produce).

\begin{figure}
\centering
\includegraphics[scale=0.26]{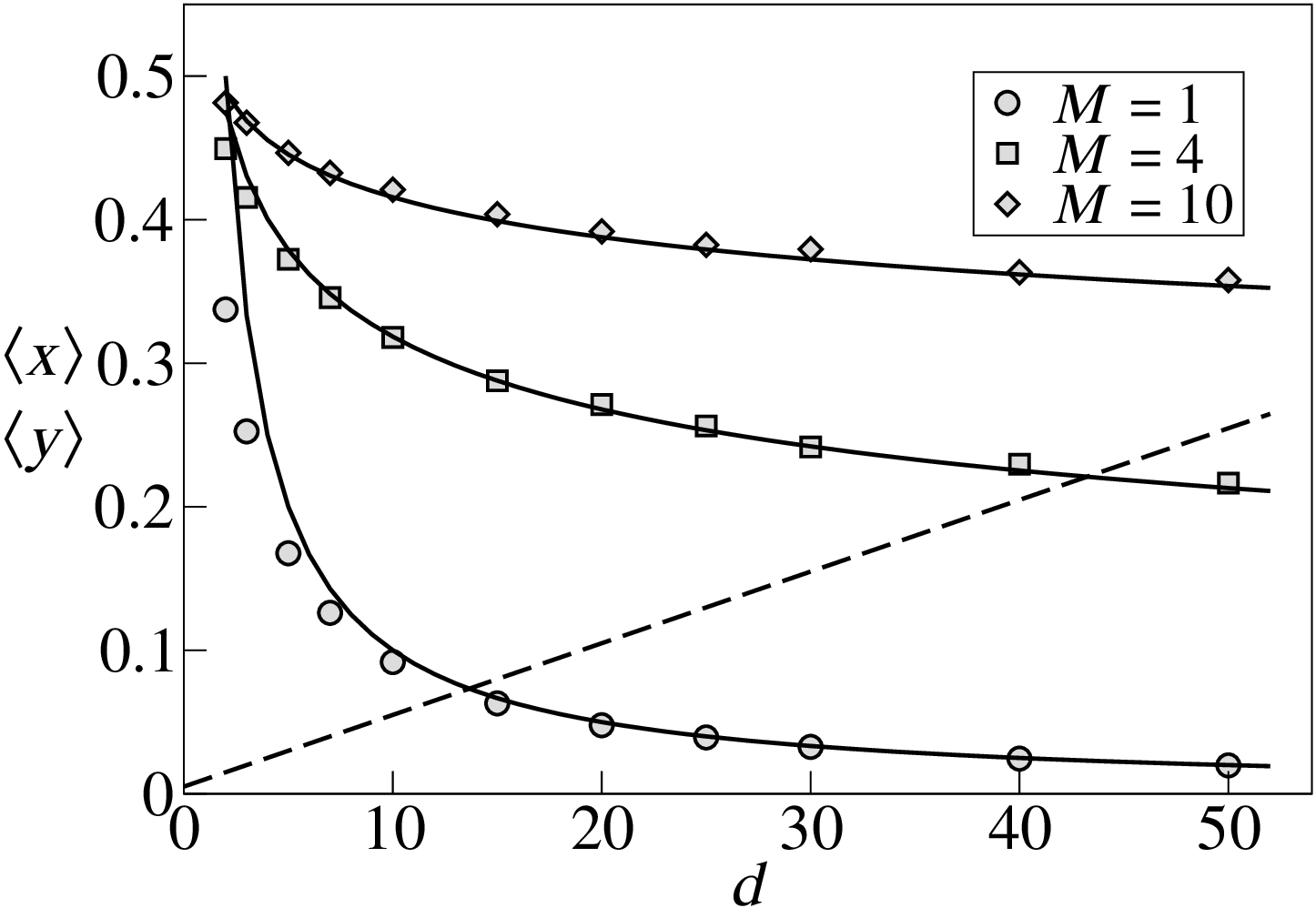}
\caption{Numerical and analytical results for $\avg{x}$ plotted
against $d$ for various values of $M$ ($N=1\,000$, numerical
results are averages of 1\,000 repetitions). The analytical
result for $\avg{y}$ is shown by the broken line.}
\label{fig-walk}
\end{figure}

Since there are no correlations in the system, when the vendor
offers his $d$ topmost variants, every variant has the same
probability to be chosen by a buyer. Thus the vendor has to go
down his list on average by $(1+d)/2$. On average this
corresponds to the cost
\begin{equation}
\label{walk-vendor}
\avg{y}=\frac{1+d}{2N}.
\end{equation}
A buyer is with the probability $1/M$ the one that has the sold
variant on the $b$th place of his list. With the complementary
probability $1-1/M$ he has this variant somewhere between the
1st and $b$th place. Thus we have
\begin{equation}
\label{walk-buyer}
\avg{x}=\frac1N\bigg(\frac 1M\,\avg{b}+\frac{M-1}{M}
\frac{1+\avg{b}}2\bigg)\approx
\frac{(M+1)\Gamma(1/M)}{2M^2}\,d^{-1/M}.
\end{equation}
When the number of buyers $M$ is large, this approaches
$\tfrac12\,d^{-1/M}$. A comparison of results
(\ref{walk-vendor}) and (\ref{walk-buyer}) with numerical
simulations is shown in fig.~\ref{fig-walk}; a good agreement is
found. A small discrepancy for $M=1$ can be corrected using the
result $\avg{y}\approx1/(d+1)$ which we develop in the following
section.

To discover the scaling behavior of $\avg{b}$, one can follow
a shorter path. The probability that one particular offered
variant is among topmost $b$ in one buyer's list is $b/N$. For
all buyers simultaneously the probability is $(b/N)^M$. Since
the vendor offers $d$ items, the probability that at least one
of them is above the line is approximately $d(b/N)^M$. If this
equals to $O(1)$, the trading is successfull. Thus we obtain
$\avg{b}=NO(1)\,d^{-1/M}$. This result scales with $M$ and $d$
in the same way as the previous outcome of the detailed
derivation.

At the end we have to mention that this model of transactions
between the vendor and buyers is not relevant for a high number
of buyers because $\avg{x}$ decreases very slowly in this case.
In other words: when $M$ is high, the probability that there is
a variant that is not worst in any buyer list approaches zero.

\subsection{A vendor producing more than one variant}
From the previous discussion we know that in a big market vendor
can not insist on selling only one variant. Therefore we would
like to investigate a more relaxed case where the vendor offers
simultaneously $d$ different variants. Then every buyer can
choose the one the most suitable for him or her. We would like
to investigate, how much suffer buyers in this case. To do so,
we label the most favorable variant from the vendor's offer for
one particular buyer $b$ and compute the average value of this
quantity.

The probability that one particular value $b$ occurs is
\begin{equation}
P(b)=\frac{d}{N-b+1}\,
\prod_{j=1}^{b-1}\bigg(1-\frac{d}{N-j+1}\bigg).
\end{equation}
In this formula the product represents the probability that all
$d$ variants offered by the vendor are in the buyer's list lower
than $b-1$, the first term represents the probability that one
of offered variants is on $b$th place in the buyer's list. Now
we can derive $\avg{b}$
\begin{equation}
\label{avg-b}
\avg{b}=\sum_{b=0}^n bP(b)=
\frac{n+1}{d+1}-d\binom{n}{d}^{-1}\approx
\frac{n+1}{d+1}.
\end{equation}
Consequently, the average cost suffered by a buyer is given by
$\avg{x}=\avg{b}/N\approx 1/(d+1)$. We see that $M$ does not
appear in $\avg{x}$. This means that the problem with the
improper behavior of the model in big markets do not appear in
this variation. Since in the calculation we did not make any
approximations, no numerical simulation is needed to check the
result.

\section{Conclusion}
The aim of this paper is to explore the modeling of a market
with inhomogeneous buyers and a vendor producing multiple
variants. We see that the outcomes depend on whether one or both
sides have adequate information about the other side or not. In
standard microeconomics, Pigou~\cite{Pigou} has introduced the
concept of price or demand elasticity. Vendors, knowing the
buyers' reserve prices to pay and thus pricing individually, can
reap significant profit---this is usually called the
first-degree price differentiation. Our analysis can be
considered as a~generalization in this direction. We show that
if individual tastes are taken into account, there is much
complexity in the system; treating individual tastes with
a~large number of buyers presents a considerable mathematical
challenge. Our models point out a convenient way to tackle this
type of problems and we expect that many real economy-motivated
problems can be analyzed in a similar way. Vendors and buyers
have many ways to improve their welfare.

In this study we have proposed two simple market models. While
accessible to analytical solutions, they exhibit many features of
real markets. In particular, diversification of the vendor's
production and market competition are used as examples. The
diversification is presented as an interplay between the
vendor's pursuit to follow the buyers' tastes and the costs
growing with the number of produced variants. We also show that
in a market with many buyers without preferences correlations,
the knowledge of these preferences doesn't increase the vendor's
profit. When correlations are introduced to the system, many
technical complications arise. Nevertheless, the results are
consistent with the expectations: a positive correlation between
the buyers' and vendor's costs improves the vendor's profit.
Also, when interests of the two parties diverge (the correlation
are negative), the vendor is able to make only a small or even
no profit. In addition, in sec.~\ref{sec-model2} a similarly
aimed model based on the well-known matching problem is
investigated.

As many other directions can be explored further, we do not
consider this topic exhausted. First of all, in a correlated
market the vendor strategies and the influence of information
deserve attention. Furthermore, while in the present work we
investigated the influence of tastes on the market, the product
quality and price were excluded from the analysis. Eventually,
the framework established herein can be used to raise the law of
supply and demand from a microscopical point of view.

\section{Acknowledgment}
We acknowledge the partial support from Swiss National Science
Foundation (project 205120-113842) as well as STIPCO (European
exchange program).

\appendix
\section{Proof of $\tau$-inequality}
\label{appendix-proof}
Let's have three lists $\vec{x},\vec{y},\vec{z}$ consisting of
$N$ mutually different real numbers. Kendall's $\tau$ for lists
$\vec{x}$ and $\vec{y}$ can be written as
$\tau_{xy}=(P_{xy}-N_{xy})/T$ where $P_{xy}$ is the number of
pairs $\alpha<\beta$ that satisfy
$(x_{\alpha}-x_{\beta})(y_{\alpha}-y_{\beta})>0$, $N_{xy}$ is
the same with a negative result of the product, and $T=N(N-1)/2$
is the total number of different pairs $\alpha,\beta$. For the
given values $\tau_{xy}$ and $N$ it follows that
\begin{equation}
\label{PaN}
P_{xy}=T(1+\tau_{xy})/2,\qquad
N_{xy}=T(1-\tau_{xy})/2.
\end{equation}

We would like to find bounds for $\tau_{yz}$ when $\tau_{xy}$
and $\tau_{xz}$ are given. First we reorder lists
$\vec{x},\vec{y},\vec{z}$ so that lists $\vec{x}$ is sorted in
the descending order and for $\alpha<\beta$ it is
$x_{\alpha}-x_{\beta}>0$. Such a rearrangement does not affect
the values of $P_{xy},P_{xz},P_{yz},N_{xy},N_{xz},N_{yz}$ and
thus the values of Kendall's tau between lists remain also
unchanged.

Since now all differences $x_{\alpha}-x_{\beta}$ are positive,
from $\tau_{xy}$ we can deduce that there are $P_{xy}$ positive
differences $y_{\alpha}-y_{\beta}$ and $N_{xy}$ negative
differences. Similarly, $P_{xz}$ differences
$z_{\alpha}-z_{\beta}$ are positive and $N_{xz}$ are negative.
The values of $P_{yz}$ and $N_{yz}$ depend on the relative
ordering of lists $\vec{y}$ and $\vec{z}$. The biggest
possible value of $P_{yz}$ occurs when positive differences
$y_{\alpha}-y_{\beta}$ are aligned with positive differences
$z_{\alpha}-z_{\beta}$ (see fig.~\ref{fig-proof}). By contrast,
the smallest value of $P_{yz}$ (and thus the smallest value
of $\tau_{yz}$) occurs when positive differences
$y_{\alpha}-y_{\beta}$ are aligned with negative differences
$z_{\alpha}-z_{\beta}$.
\begin{figure}
\centering
\includegraphics[scale=1]{buysell2}\\[4mm]
\includegraphics[scale=1]{buysell3}
\caption{An illustration of the proof. The first case (first
three lines) has the biggest possible value of $P_{yz}$, the
second case has the smallest possible value of $P_{yz}$.}
\label{fig-proof}
\end{figure}

From fig.~\ref{fig-proof} we see that $P_{yz}$ and $N_{yz}$
fulfill the inequalities
\begin{equation*}
N_{yz}\geq\abs{P_{xy}-N_{xz}},\qquad
P_{yz}\geq\abs{P_{xy}-P_{xz}}.
\end{equation*}
Using $P_{ab}+N_{ab}=T$ and (\ref{PaN}) we obtain
\begin{align*}
\tfrac T2\abs{\tau_{xy}-\tau_{yz}}\leq&P_{yz}\leq
T-\tfrac T2\abs{\tau_{xy}+\tau_{yz}},\\
-T+\tfrac T2\abs{\tau_{xy}-\tau_{yz}}\leq&-M_{yz}\leq
-\tfrac T2\abs{\tau_{xy}+\tau_{yz}}.
\end{align*}
These two inequalities summed together and divided by $T$
yield the desired inequality (\ref{tauinequality}).

\section{Expected values of $\avg{\tau}$}
\label{appendix-tau}
For lists created using (\ref{lists}) we can rearrange
(\ref{kandall}) as follows
\begin{equation*}
\avg{\tau_{xx}}=\frac2{N(N-1)}
\sum_{\alpha<\beta}\avg{\sigma_{\alpha\beta}}=
\avg{\sigma_{\alpha\beta}}.
\end{equation*}
Moreover, $\sigma_{\alpha\beta}$ can be rewritten as
\begin{equation*}
\avg{\sigma_{\alpha\beta}}=P_{++}+P_{--}-P_{-+}-P_{+-}=
1-2P_{-+}-2P_{+-}=1-4P_{+-}.
\end{equation*}
Here $P_{++}$ is the probability that both
$x_{\alpha}-x_{\beta}$ and $x_{\alpha}'-x_{\beta}'$ are positive
and so forth, the formulae $P_{+-}=P_{-+}$,
$P_{++}+P_{--}+P_{-+}+P_{+-}=1$ are used. According to
(\ref{lists}) we write
\begin{align*}
x_{\alpha}-x_{\beta}&=(1-t)(a_{\alpha}-a_{\beta})+
t(c_{\alpha}-c_{\beta})\equiv (1-t)A+tC,\\
x_{\alpha}'-x_{\beta}'&=(1-t)(b_{\alpha}-b_{\beta})+
t(c_{\alpha}-c_{\beta})\equiv (1-t)B+tC
\end{align*}
where $A,B,C$ lie in the range $[-1;1]$ and are equally
distributed with the density $\varrho(A)=1-\abs{A}$.
Now we have ($t/(1-t)\equiv u$)
\begin{equation*}
P(+-\vert C)=\begin{cases}
\tfrac12(uC)^2\big[1-\tfrac12(uC)^2\big] & (C\leq1/u),\\
0 & (C>1/u).
\end{cases}
\end{equation*}
If $u\leq1$, the first case applies to all possible values of
$C$, $P(+-\vert C)=0$ is possible only if $u>1$. Finally, using
\begin{equation*}
P(+-)=\int_{-1}^1P(+-\vert C)\varrho(C)\,\dd C
\end{equation*}
with $\varrho(C)=1-\abs{C}$ it follows that
\begin{equation*}
\avg{\tau_{xx}}=\begin{cases}
\frac{u^2}{15}\big(10-6u+u^2\big) & (u\leq1),\\
\frac1{15}\big(15-\tfrac{14}u+\tfrac4{u^2}\big) & (u>1).
\end{cases}
\end{equation*}
The quantity $\avg{\tau_{xy}}$ can be derived in the same way.

The variance of $\tau_{xx}$ can be found by a direct computation
of $\avg{\tau_{xx}^2}$. We have
\begin{equation*}
\tau_{xx}^2=\frac1{N^2(N-1)^2}\bigg(
\sum_{\alpha\neq\beta}\sigma_{\alpha\beta}^2+
\sum_{\alpha\neq\beta}\sum_{\gamma\neq\delta}
\sigma_{\alpha\beta}\sigma_{\gamma\delta}\bigg).
\end{equation*}
The averaging procedure is straightforward. At the end we
obtain
\begin{equation*}
\sigma_{\tau}^2=\avg{\tau_{xx}^2}-\avg{\tau_{xx}}^2\approx
\frac4N\,\Big(\avg{\sigma_{\alpha\gamma}\sigma_{\gamma\beta}}-
\avg{\sigma_{\alpha\beta}}^2\Big),
\end{equation*}
where terms proportional to higher powers of $1/N$ were
neglected. The variance is largest when $t=0$, for $t=\pm1$
obviously $\sigma_{\tau}=0$.

\section{System's upper bound with a given $\tau_0$}
\label{appendix-ubound}
To obtain the upper limit $M_m$ for the number of lists that
have pairwise Kendall's tau equal to $\tau_0$ we use a
constructive way of reasoning. Without a loss of generality we
assume that in the first list all $N(N-1)/2\equiv T$ differences
$x_{\alpha}-x_{\beta}$ ($\alpha<\beta$) are positive. From
eq.~(\ref{PaN}) it follows that to maintain $\tau_{12}=\tau_0$,
in the second list exactly $N_2=(1-\tau_0)T/2$ pairs have to be
negative. The same holds for all other lists we would like to
add to the system, $N_i=(1-\tau_0)T/2\equiv W$ is required for
$i\geq2$ ($N_1=0$ is fixed by the chosen ordering of the first
list).

\begin{figure}
\centering
\includegraphics[scale=1]{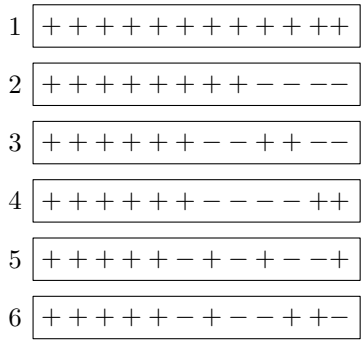}
\caption{Construction of the set of lists with the given value
of Kendall's tau for $T=10$ (thus $N=5$) and $W=4$ (thus
$\tau_0=0.2$).}
\label{fig-construction}
\end{figure}

To represent the way of construction we use
fig.~\ref{fig-construction}. The first list is represented by
$T$ positive pairs there, second one by $T-N_2$ positive pairs
and by $N_2$ negative pairs. We would like to find a third list
satisfying $\tau_{13}=\tau_{23}=\tau_0$. We already now that
$N_3=W$ negative pairs have to be accommodated there. To achieve
$\tau_{23}=\tau_0$ it is necessary that exactly half of the
negative pairs in the second list meets with negative pairs in
the third list. This can be fulfilled in two ways (lines 3 and 4
in fig.~\ref{fig-construction}) which also have mutually
$\tau_{34}=\tau_0$.

Now there are two important points to notice. First, the
construction of the third and the fourth list is impossible when
$N_3$ is an odd number. Thus for corresponding values of
$\tau_0$ there are not more than two lists that have mutually
Kendall's tau equal to this $\tau_0$. Second, we have used
purely combinatorial arguments here without taking care whether
described set of lists (e.g. lists 1, 2, 3 and 4 from
fig.~\ref{fig-construction}) do exist. Thus we estimate an upper
bound which cannot be exceeded but which can dwarf the real
maximum by far.

Now we can continue with the fifth list where again $N_5=W$
negative pairs are present. Among them exactly $N_5/2$ have to
meet with negative pairs in the second list and also half of
them have to meet with negative pairs in the third and fourth
list. This can be achieved by lists 5 and 6 in
fig.~\ref{fig-construction}), their relative Kendall's tau is
also $\tau_0$.

During the construction process we divide $W$ negative pairs
present in the second lists into two groups with size $W/2$
(lists 3 and 4), then we divide again and obtain lists 5 and 6
with groups of negative pairs with size $W/4$. Clearly, this
sequence ends when we divide the original number of negative
pairs $W$ so many times that we arrive at $1$ which cannot be
divided further. Consequently, the upper bound we are looking
for is the following
\begin{equation}
\label{M_m1}
M_m\approx 2+2\log_2 w=2+2\log_2\frac{(1-\tau_0)(N-1)N}4.
\end{equation}
This formula holds when $\tau_0<1$ (for $\tau_0=1$, clearly
$M_m=1$).

In the previous construction, there is one hidden flaw. Since
only $W/2$ negative pairs in third list meet negative pairs in
second list, to prepare lists 1, 2 and 3 we use altogether
$W+W/2$ pairs. If this number is greater then $T$, the third
list cannot be constructed in the described way and the limiting
$M_m$ differs from the previously found result. To construct
lists 5 and 6 we need together $W+W/2+W/4$ pairs which has to be
smaller then $T$. By generalizing previous argument we can write
down the inequality for the maximum number of lists $M_m$
\begin{equation*}
W+W/2+\dots+W/2^{M_m/2-1}\leq T.
\end{equation*}
For $M_m$ itself it follows that
\begin{equation}
\label{M_m2}
M_m\leq 2\log_2\frac{1-\tau_0}{-\tau_0}.
\end{equation}
This limit is relevant only for $\tau_0<0$ (with $\tau_0>0$ we
have $W<T/2$ and $W+W/2+\dots$ is less then $T$). As an actual
upper bound, the smaller value from (\ref{M_m1}) and
(\ref{M_m2}) applies.

\end{document}